\documentclass[twocolumn]{aastex63}
\usepackage{amsmath}
\usepackage{color}
\usepackage{natbib}
\usepackage{enumitem}
\usepackage{hyperref}
\usepackage{listings}
\usepackage{fontawesome}
\usepackage{xcolor}
\usepackage{xspace} 
\usepackage{url}
\usepackage{fontawesome}
\usepackage{graphicx,subfigure}

\usepackage{lineno}
\newcommand{\comment}[1]{{#1}}

\definecolor{linkcolor}{rgb}{0.,0.3,0.7}
\definecolor{codegreen}{rgb}{0,0.6,0}
\definecolor{codegray}{rgb}{0.5,0.5,0.5}
\definecolor{codepurple}{rgb}{0.58,0,0.82}
\definecolor{backcolour}{rgb}{0.95,0.95,0.92}

\lstdefinestyle{mystyle}{
    commentstyle=\color{codegray},
    keywordstyle=\color{codegreen},
    numberstyle=\tiny\color{gray},
    stringstyle=\color{codepurple},
    basicstyle=\ttfamily\footnotesize,
    breakatwhitespace=false,         
    breaklines=true,                 
    captionpos=b,                    
    keepspaces=true,                 
    numbers=none,                    
    numbersep=5pt,                  
    showspaces=false,                
    showstringspaces=false,
    showtabs=false,                  
    tabsize=2
}

\hypersetup{breaklinks, colorlinks,
  urlcolor=linkcolor,linkcolor=linkcolor,citecolor=linkcolor}
\newcommand{\Angs}{\ensuremath{\text{\AA}}}

\newcommand{\Ha}{H$\rm\alpha$}
\newcommand{\Oiii}{[\mbox{O\,{\sc iii}}]}
\newcommand{\Oii}{[\mbox{O\,{\sc ii}}]}
\newcommand{\Lya}{Ly$\rm\alpha$}

\submitjournal{ApJ}

\graphicspath{{figures/}}
 
\shorttitle{The Merian Survey}

\begin{document}\sloppy\sloppypar\raggedbottom\frenchspacing 

\title{Merian: A Wide-Field Imaging Survey of Dwarf Galaxies at $z \sim 0.06-0.10$}

\author[0000-0002-1841-2252]{Shany Danieli}
\affil{Department of Astrophysical Sciences, Princeton University, 4 Ivy Lane, Princeton, NJ 08544, USA}
\affil{School of Physics and Astronomy, Tel Aviv University, Tel Aviv 69978, Israel}

\author[0000-0002-0332-177X]{Erin Kado-Fong}
\affil{Physics Department, Yale Center for Astronomy \& Astrophysics, PO Box 208120, New Haven, CT 06520, USA}

\author[0000-0003-1385-7591]{Song Huang}
\affil{Department of Astronomy, Tsinghua University, Beijing 100084, China}

\author[0000-0001-7729-6629]{Yifei Luo}
\affil{Department of Astronomy and Astrophysics, University of California, Santa Cruz, 1156 High Street, Santa Cruz, CA 95064 USA}

\author[0000-0002-9110-6163]{Ting S Li}
\affil{Department of Astronomy and Astrophysics, University of Toronto, 50 St. George Street, Toronto ON, M5S 3H4, Canada}

\author[0000-0001-9395-4759]{Lee S Kelvin}
\affil{Department of Astrophysical Sciences, Princeton University, 4 Ivy Lane, Princeton, NJ 08544, USA}

\author[0000-0002-3677-3617]{Alexie Leauthaud}
\affil{Department of Astronomy and Astrophysics, University of California, Santa Cruz, 1156 High Street, Santa Cruz, CA 95064 USA}

\author[0000-0002-5612-3427]{Jenny E. Greene}
\affil{Department of Astrophysical Sciences, Princeton University, 4 Ivy Lane, Princeton, NJ 08544, USA}

\author[0000-0002-9816-9300]{Abby Mintz}
\affil{Department of Astrophysical Sciences, Princeton University, 4 Ivy Lane, Princeton, NJ 08544, USA}

\author[0000-0001-6052-4234]{Xiaojing Lin}
\affil{Department of Astronomy, Tsinghua University, Beijing 100084, China}
\affil{Steward Observatory, University of Arizona, 933 N Cherry Ave, Tucson, AZ 85721, USA}

\author[0000-0001-9592-4190]{Jiaxuan Li}
\affil{Department of Astrophysical Sciences, Princeton University, 4 Ivy Lane, Princeton, NJ 08544, USA}

\author[0000-0003-4703-7276]{Vivienne Baldassare}
\affil{Department of Physics and Astronomy, Washington State University, Pullman, WA 99163, USA}

\author[0000-0002-5209-1173]{Arka Banerjee}
\affil{Department of Physics, Indian Institute of Science Education and Research}

\author[0000-0001-6442-5786]{Joy Bhattacharyya}
\affil{Department of Astronomy, The Ohio State University, Columbus, OH 43210, USA}
\affil{Center for Cosmology and Astro-Particle Physics, The Ohio State University, Columbus, OH 43210, USA}

\author{Diana Blanco}
\affil{Department of Astronomy and Astrophysics, University of California, Santa Cruz, 1156 High Street, Santa Cruz, CA 95064 USA}

\author[0000-0002-0372-3736]{Alyson Brooks}
\affiliation{Department of Physics and Astronomy, Rutgers, The State University of New Jersey, 136 Frelinghuysen Rd, Piscataway, NJ 08854, USA}
\affiliation{Center for Computational Astrophysics, Flatiron Institute, 162 Fifth Ave, New York, NY 10010, USA}

\author[0000-0001-8467-6478]{Zheng Cai}
\affil{Department of Astronomy, Tsinghua University, Beijing 100084, China}

\author{Xinjun Chen}
\affil{Department of Astronomy and Astrophysics, University of California, Santa Cruz, 1156 High Street, Santa Cruz, CA 95064 USA}

\author[0000-0001-7831-4892]{Akaxia Cruz}
\affil{Center for Computational Astrophysics, Flatiron Institute, 162 Fifth Ave, New York, NY 10010, USA}
\affil{Department of Astrophysical Sciences, Princeton University, 4 Ivy Lane, Princeton, NJ 08544, USA}
\affil{Department of Physics, Princeton University, 4 Ivy Lane, Princeton, NJ 08544, USA}

\author[0000-0003-1509-9966]{Robel Geda}
\affil{Department of Astrophysical Sciences, Princeton University, 4 Ivy Lane, Princeton, NJ 08544, USA}

\author{Runquan Guan}
\affil{Department of Astronomy and Astrophysics, University of California, Santa Cruz, 1156 High Street, Santa Cruz, CA 95064 USA}

\author[0000-0001-9487-8583]{Sean Johnson}
\affil{Department of Astronomy, University of Michigan, Ann Arbor, MI 48109, USA}

\author[0000-0001-8783-6529]{Arun Kannawadi}
\affil{Department of Astrophysical Sciences, Princeton University, 4 Ivy Lane, Princeton, NJ 08544, USA}
\affil{Department of Physics, Duke University, Box 90305, Durham, NC 27708, USA}

\author[0000-0001-7052-6647]{Stacy Y. Kim}
\affil{Carnegie Observatories, 813 Santa Barbara St, Pasadena, CA 91101, USA}

\author[0000-0001-6251-649X]{Mingyu Li}
\affil{Department of Astronomy, Tsinghua University, Beijing 100084, China}

\author[0000-0003-1666-0962]{Robert Lupton}
\affil{Department of Astrophysical Sciences, Princeton University, 4 Ivy Lane, Princeton, NJ 08544, USA}

\author[0000-0002-9419-6547]{Charlie Mace}
\affil{Department of Physics, The Ohio State University, Columbus, OH 43210, USA}
\affil{Center for Cosmology and Astro-Particle Physics, The Ohio State University, Columbus, OH 43210, USA}

\author[0000-0003-0105-9576]{Gustavo E. Medina}
\affiliation{Department of Astronomy \& Astrophysics, University of Toronto, Toronto, ON M5S 3H4, Canada}

\author[0000-0002-7922-9726]{Yue Pan}
\affil{Department of Astrophysical Sciences, Princeton University, 4 Ivy Lane, Princeton, NJ 08544, USA}

\author[0000-0002-8040-6785]{Annika H. G. Peter}
\affil{Department of Physics, The Ohio State University, Columbus, OH 43210, USA}
\affil{Department of Astronomy, The Ohio State University, Columbus, OH 43210, USA}
\affil{Center for Cosmology and Astro-Particle Physics, The Ohio State University, Columbus, OH 43210, USA}

\author[0000-0002-1164-9302]{Justin I. Read}
\affil{Department of Physics, University of Surrey, Guildford, GU2 7XH, United Kingdom}

\author[0000-0002-7967-7676]{Rodrigo Córdova Rosado}
\affil{Department of Astrophysical Sciences, Princeton University, 4 Ivy Lane, Princeton, NJ 08544, USA}

\author{Allen Seifert}
\affil{Department of Astronomy and Astrophysics, University of California, Santa Cruz, 1156 High Street, Santa Cruz, CA 95064 USA}

\author[0000-0003-3986-9427]{Erik J. Wasleske}
\affil{Department of Physics and Astronomy, Washington State University, Pullman, WA 99163, USA}

\author[0000-0001-9833-6183]{Joseph Wick}
\affil{Department of Astronomy and Astrophysics, University of California, Santa Cruz, 1156 High Street, Santa Cruz, CA 95064 USA}

\begin{abstract}

We present the Merian Survey, an optical imaging survey optimized for studying the physical properties of bright star-forming dwarf galaxies. Merian is carried out with two medium-band filters ($N708$ and $N540$, centered at $708$ and $540$ nm), custom-built for the Dark Energy Camera (DECam) on the Blanco telescope. Merian covers $\sim 750\,\mathrm{deg}^2$ of equatorial fields, overlapping with the Hyper Suprime-Cam Subaru Strategic Program (HSC-SSP) wide, deep, and ultra-deep fields. When combined with the HSC-SSP imaging data ($grizy$), the new Merian DECam medium-band imaging allows for photometric redshift measurements via the detection of \Ha\ and \Oiii\ line emission flux excess in the $N708$ and $N540$ filters, respectively, at $0.06<z<0.10$. \comment{We present an overview of the survey design, observations taken to date, data reduction using the LSST Science Pipelines, including aperture-matched photometry for accurate galaxy colors, and a description of the data included in the first data release (DR1).} The key science goals of Merian include: probing the dark matter halos of dwarf galaxies out to their virial radii using high signal-to-noise weak lensing profile measurements, decoupling the effects of baryonic processes from dark matter, and understanding the role of black holes in dwarf galaxy evolution. This rich dataset will also offer unique opportunities for studying extremely metal-poor galaxies via their strong \Oiii\ emission and \Ha\ lines, as well as \Oii\ emitters at $z\sim 0.4$, and \Lya\ emitters at $z\sim 3.3$ and $z\sim 4.8$. Merian showcases the power of utilizing narrow and medium-band filters alongside broad-band filters for sky imaging, demonstrating their synergistic capacity to unveil astrophysical insights across diverse astrophysical phenomena. 

\end{abstract}


\section{Introduction} \label{sec:intro}

\begin{figure*}[t]
 \centering
 \includegraphics[width=1.0\textwidth]{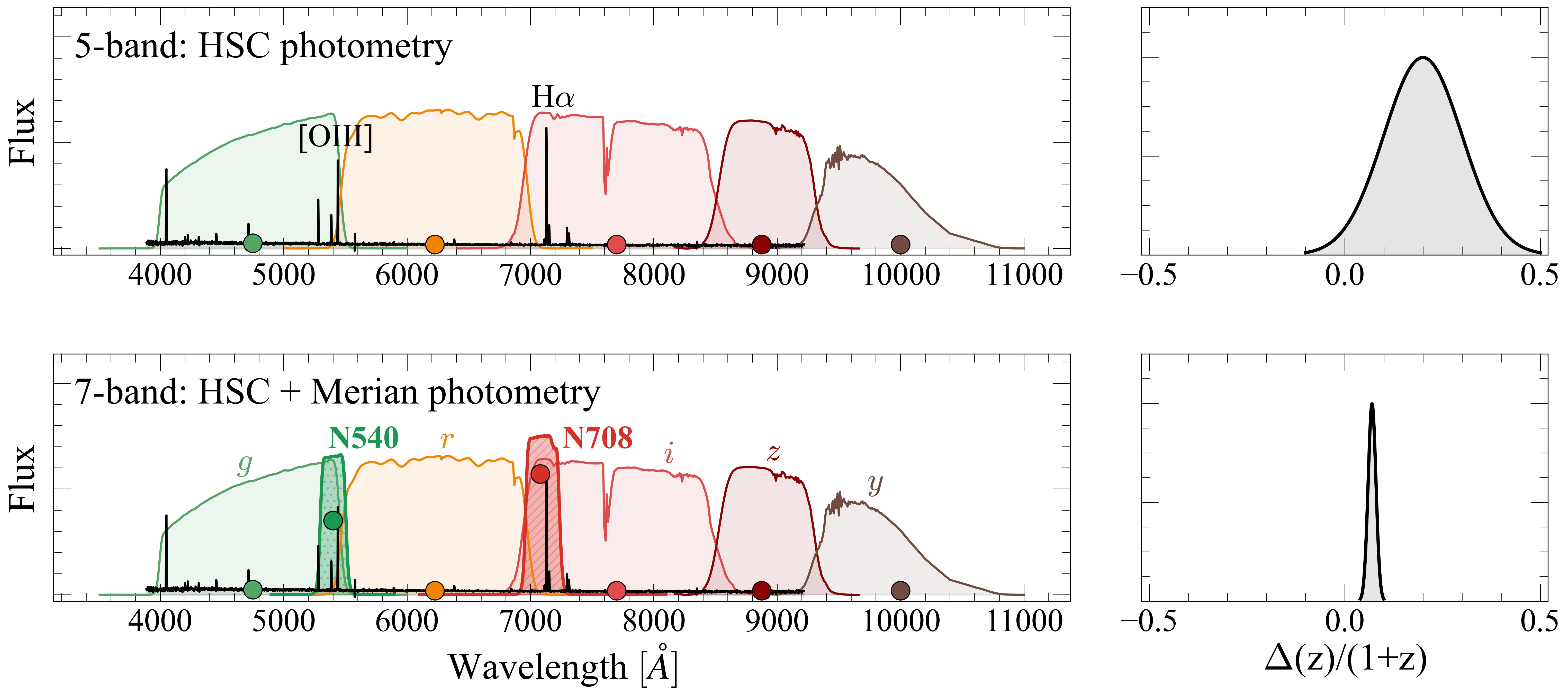}
  \caption{Left panels: Spectrum of a star-forming dwarf galaxy at $z=0.086$ from SDSS (black), overlaid with the HSC broad bands (upper left panel) and with the addition of the Merian medium bands (lower left panel). Photometry from the two Merian medium band filters, $N708$ and $N540$ (\citealt{Luo:2023}), when added to the photometry from the five HSC broad bands, $grizy$, provides a significant improvement in our ability to recover the redshift of dwarfs. The right panels show the redshift accuracy when using just the five-band HSC photometry (upper right panel) compared to the full seven-band photometry when the two Merian bands are added (lower right panel).
  }
  \label{fig:motivation_fig}
\end{figure*}

Ever since the commissioning of the Sloan Digital Sky Survey \citep[SDSS;][]{Gunn:1998}, digital wide-field imaging surveys have been instrumental in advancing various astrophysical domains \citep[e.g.,][]{Gwyn:2012, deJong:2013, Chambers:2016, DES:2016}. Such surveys have greatly enriched our understanding of the universe by generating detailed sky maps that sample large numbers of astronomical objects. They have contributed to a wide range of astronomical studies, from the origins of our universe through the formation of extrasolar planets. Now, we are embarking on the next generation of wide-field imaging surveys, utilizing increasingly larger telescopes to explore ever-expanding areas of the sky with unprecedented detail and precision. The Rubin Observatory \citep{Ivezic:2019} will soon commence its ten-year Legacy Survey of Space and Time (LSST), conducting an extensive and deep survey over a vast expanse of the sky using the 8.4-meter Simonyi Survey Telescope.

Many wide-field imaging surveys have traditionally relied on broad-band filters. However, integrating narrow and medium-band imaging alongside these surveys is increasingly recognized for its potential benefits, as recently demonstrated by several pioneering surveys. Narrow-band filters, for instance, enable precise observations of specific emission lines from galaxies, shedding light on their ionization states \citep[e.g.,][]{Lee:2009, Aihara:2018, Ouchi:2018} and star formation fueling \citep[e.g.,][]{Lokhorst:2022}. They are also utilized in chemical composition studies via their stellar absorption features \citep[e.g.,][]{Stromgren:1966, Eggen:1976, Richtler:1989, Starkenburg:2017, Chiti:2020, Fu:2023}. Meanwhile, medium-band filters offer enhanced spectral resolution, allowing for more accurate photometric redshift measurements and detailed characterization of distant galaxies \citep[e.g.,][]{Wolf:2003, Ilbert:2009, Whitaker:2011, Padilla:2019}. Nevertheless, a persistent challenge remains the limited sky coverage typically associated with traditional narrow and medium-band imaging surveys.

\comment{Here, we introduce the Merian Survey, a novel medium-band survey to augment the deep, wide-field multi-broad-band ($grizy$) imaging from the Hyper Suprime-Cam Subaru Strategic Program (HSC-SSP), a public imaging dataset obtained with the 8.2-meter Subaru telescope \citep{Aihara:2018}. The Merian Survey adds new imaging of the same area using two custom-made medium-band filters \citep{Luo:2023} mounted on the Blanco/Dark Energy Camera (DECam; \citealt{Flaugher:2015}).} Merian addresses a limitation of current wide-field spectroscopic surveys by enabling the acquisition of a large statistical sample of bright star-forming dwarf galaxies with accurate photometric redshifts (Figure \ref{fig:motivation_fig}) and spatially resolved information (Figures \ref{fig:example_single} and \ref{fig:gallery}). Its primary scientific objectives include measuring halo masses of dwarfs through galaxy-galaxy lensing and clustering \citep{Leauthaud:2020}, as well as obtaining detailed measurements of dwarf galaxy structures and star formation rates in a mass range where feedback and other baryonic processes play crucial but poorly understood roles. The innovative approach has the potential to advance understanding of a broad range of astrophysical phenomena. 

The technical lessons learned from the Merian Survey will be readily transferrable to future wide-field imaging surveys like LSST, Roman \citep{Green:2012, Spergel:2015, Akeson:2019}, and Euclid \citep{Laureijs:2011, Euclid:2024}. These lessons include strategies for combining data from different telescopes and the demonstrated power of medium-band filters in large-scale imaging surveys. By paving the way for such endeavors, we hope that Merian could serve not only as a significant scientific effort in its own right but also as a crucial precursor to future advancements in observational astronomy.

An outline of the paper is as follows. The paper begins with a description of the principles guiding the survey design, particularly the Merian filters' characteristics, the survey fields, and the required depth (Section \ref{sec:survey_design}). \comment{In Section \ref{sec:obs}, we introduce the observation strategy, present the first data release (DR1) of the survey, and provide an overview of the observations}. This is followed by descriptions of the data reduction in Section \ref{sec:data-reduction} and the photometric catalogs in Section \ref{sec:photometry-catalog}. In Section \ref{sec:science_objs} we outline our key science objectives and we summarize in Section \ref{sec:summary}.

\begin{figure*}[t]
 \centering
 \includegraphics[width=1.0\textwidth]{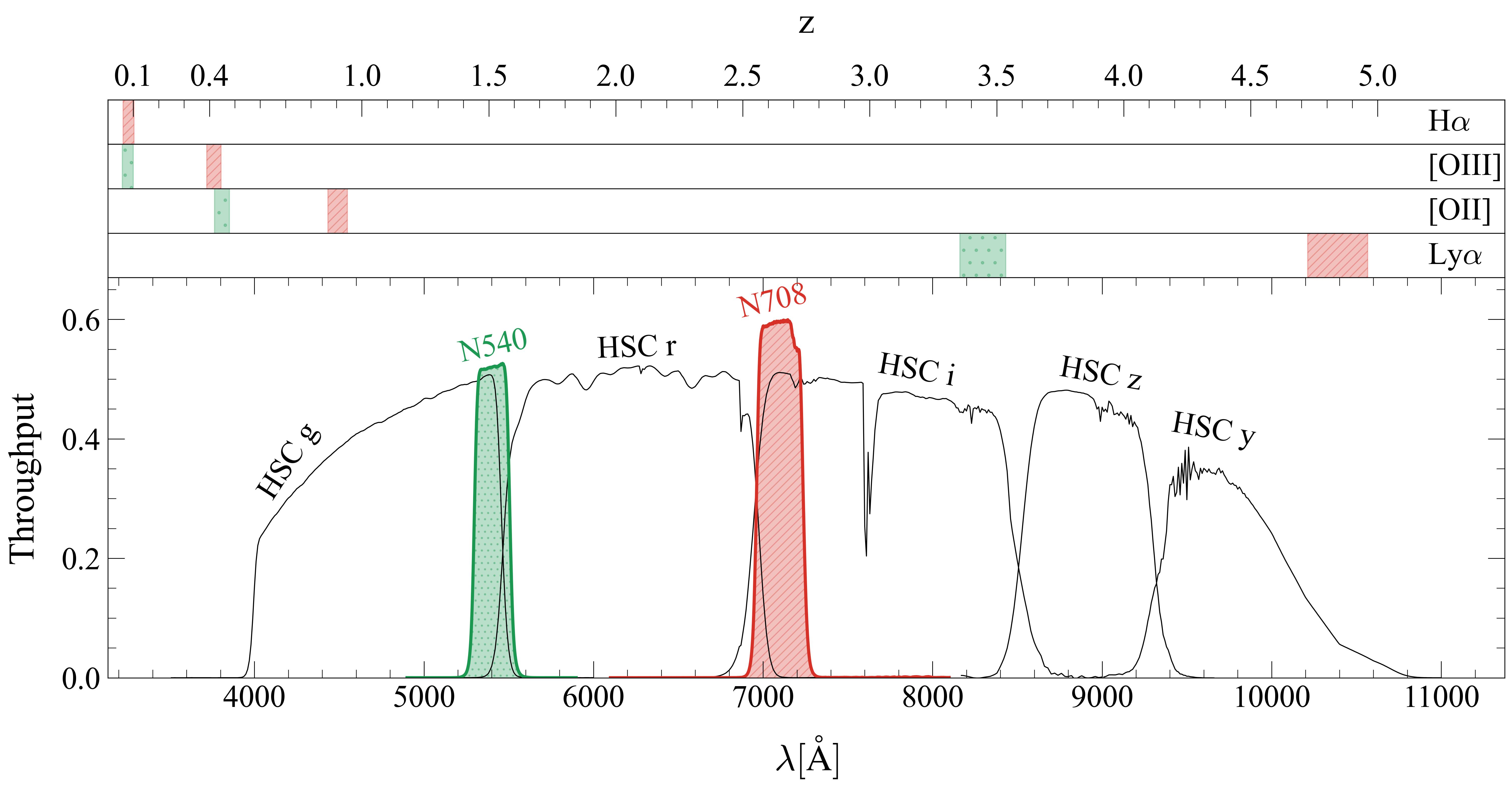}
  \caption{Throughput curves of the $N540$ (green) and $N708$ (red) medium-band filters customized for the Merian survey \citep{Luo:2023}, alongside the HSC $grizy$ broad-band filters (black) used in the HSC-SSP survey \citep{Aihara:2018}. The shaded bands in the top panel indicate the redshift range in which rest-frame optical spectral lines of \Ha, \Oiii, \Oii\, and \Lya\ fall within the two Merian filters.}
  \label{fig:trans_curves}
\end{figure*}

\section{Survey design}\label{sec:survey_design}

\subsection{Survey Design Principles}
The Merian survey combines the publicly available dataset from the HSC-SSP survey, a wide-field broad-band ($grizy$; Figures \ref{fig:motivation_fig},\ref{fig:trans_curves}) imaging survey with the Subaru \comment{8.2-meter} telescope \citep{Aihara:2018}, with dedicated imaging taken with DECam on the Victor M. Blanco 4-m Telescope using a set of custom-designed medium-band filters, $N708$, and $N540$, to cover the \Ha $/6563\Angs$ and \Oiii $/5007\Angs$ lines at $0.06<z<0.10$. This section outlines the criteria guiding the survey design regarding the selected redshift range, area, and target stellar mass range.

\vspace{0.1cm}
\noindent
\textbf{Sensitivity.} 
The depth of the HSC-SSP broad-band images sets the lowest achievable limiting stellar mass. Based on studies of typical dwarfs in the Local Volume ($D<10\,\mathrm{Mpc}$), a stellar mass of $M_\star=10^8\,M_\odot$ corresponds to an effective surface brightness brighter than $\mu_{\mathrm{eff},V}=26\,\mathrm{mag\,arcsec^{-2}}$ \citep[e.g.,][]{Danieli:2018, Carlsten:2021b, Carlsten:2022b}. The \texttt{Wide} layer of the HSC-SSP covers $1100\,\mathrm{deg^2}$ and based on the third data release from this program (PDR3; \citealt{Aihara:2022}), it reaches a full depth of $m_r\sim 26\,\mathrm{mag}$ at $5\sigma$ in all five broad-band filters ($grizy$). \citet{Li:2023a} performed a large suite of image simulations to derive the completeness of the HSC-SSP \texttt{Wide} layer images from the Public Data Release 2 (PDR2, also known as \texttt{S18A}; \citealt{Aihara:2018}). They find $>70\%$ completeness for galaxies with $\mu_{\mathrm{eff},g}<26.5\,\mathrm{mag\,arcsec^{-2}}$ (see also \citealt{Greco:2018}). By performing similar fake galaxy injection simulations, \citet{Leauthaud:2020} estimated that the HSC-SSP \texttt{Wide} layer is mass complete to $\log (M_{\star,\mathrm{lim}}/M_\odot)=7.3$ out to $z=0.1$ and to $\log (M_{\star,\mathrm{lim}}/M_\odot)=8.1$ out to $z=0.3$.

\vspace{0.1cm}
\noindent
\textbf{Sample Size/Volume.} The scientific specification that places the most stringent requirement on the Merian sample size is the number of dwarfs needed to measure the lensing signal via weak (galaxy-galaxy) gravitational lensing. \citet{Leauthaud:2020} computed the predicted amplitude of the galaxy-galaxy lensing ($\Delta \Sigma$) for dwarfs in the HSC-SSP \texttt{Wide} layer, assuming an observed area of $1000\,\mathrm{deg^2}$, within the redshift range $0<z<0.25$. For galaxies in two narrow ($\Delta M_\star=0.2\,\mathrm{dex}$) mass bins centered around $\log (M_\star/M_\odot)=8$ and $\log (M_\star/M_\odot)=9$, the predicted S/N at $r<500\,\mathrm{kpc}$ is 37 and 46, respectively. In the one-halo regime ($R_\mathrm{200m}$), the predicted S/N for $\log (M_\star/M_\odot)=8$ and $\log (M_\star/M_\odot)=9$ is 8 and 15. Thus, measurements of the one and two-halo terms for dwarf galaxies could be obtained by surveying an equivalent volume with HSC-SSP. The combination of sensitivity and volume, as required by the lensing S/N, will also support our secondary objective of comprehensively exploring dwarfs' key properties, encompassing stellar masses, sizes, star formation rates, down to $M_\star\sim 10^8\,M_\odot$. By sampling the parameter space of these properties both extensively (thanks to the large sample size) and in a well-described manner (through medium-band selection on line emission strength), our approach meets the technical specifications needed to measure the lensing sample from dwarf stellar masses while also providing a large, sensitive, and spatially resolved sample of line emitters in our redshift range. Specifically, it enables the exploration of covariance among these key parameters while controlling for stellar mass.

\vspace{0.1cm}
\noindent
\textbf{Redshifts.} Accurate (but not necessarily precise) galaxy redshifts are key to our broad Merian science goals in lensing, baryonic processes, and beyond. In \citet{Luo:2023}, we inferred the photometric redshifts of galaxies in the COSMOS field ($m_r<24\,\mathrm{mag}$) using the $grizy$ HSC-SSP photometry alone and compared them to the photometric redshifts obtained using the 30-band COSMOS catalog \citep{Laigle:2016}. At $z<0.1$, the five-band HSC-SSP photometry yields a redshift accuracy of $\sigma_{\Delta z/(1+z)}\sim 0.5$ and a completeness and purity of $48\%$ and $12\%$, respectively, that is unacceptable for most science cases. As shown in \citet{Luo:2023} using image simulations, adding the $N708$ and $N540$ photometry improves the photometric redshift accuracy to $\sigma_{\Delta z/(1+z)}\sim 0.01$, with $89\%$ completeness and $90\%$ purity at $z<0.1$. \comment{Figure \ref{fig:motivation_fig} demonstrates the power of incorporating the Merian medium-band filters alongside the five HSC-SSP broad bands, compared to using the broad-band photometry alone. The $N540$ and $N708$ filters capture the \Oiii\ and \Ha\ emission lines, respectively. Adding these medium-band filters enhances the redshift accuracy by an order of magnitude to $\sigma_{\Delta z/(1+z)}\sim 1-2\%$, enabling substantially more precise measurements of galaxy properties and their environments.} This redshift accuracy and precision level is sufficient for achieving our science goals, ensuring reliable measurements for our primary objectives.

\subsection{Filter Design and Characteristics}

We introduced the custom-designed filter set for the DECam on the 4-meter Victor M. Blanco telescope at the Cerro Tololo Inter-American Observatory in a dedicated Merian filter design paper \citep{Luo:2023}. We briefly recount the design specifications here for the reader's convenience and completeness. 

The dual filter system comprises two medium-band filters: the $N540$ filter centered at $\lambda_\mathrm{c} = 5400\Angs$ and $\Delta\lambda = 210\Angs$, and the $N708$ filter at $\lambda_\mathrm{c}=7080\Angs$ with a width of $\Delta\lambda=275\Angs$, and , as shown in Figure \ref{fig:trans_curves}. The filters detect rest-frame \Oiii\ and \Ha\ emission, respectively, from galaxies at a redshift window of $0.06<z<0.10$. The central wavelength and bandwidth were tuned for Merian's flagship weak lensing analysis for dwarf galaxies. The primary objective of the dual filter approach is to measure photometric redshifts and remove high redshift interlopers. Accordingly, three separate requirements were taken into account when designing the filters: (1) Achieving a low outlier fraction ($\eta$), driven by high sample completeness and the accuracy of the photometric redshifts; (2) Attaining a high S/N measurement of the galaxy-galaxy lensing signal, which directly depends on the number of lens dwarf galaxies; (3) Avoiding strong skylines. 

To balance all of these requirements, in \citet{Luo:2023}, we performed image simulations to assess a wide set of potential filter designs for the \Ha\ filter, characterized by the central wavelength, $\lambda_\mathrm{c}$ and filter bandwidth, $\Delta\lambda$. These simulations were constructed to predict the ability of the filters to detect \Ha\ and \Oiii\ emission lines from bright dwarfs, and to quantify the survey's expected photometric redshift accuracy and precision. In short, the full survey was forward-modeled for a range of [$\lambda_\mathrm{c}$,$\Delta\lambda$] pairs, spanning a filter central wavelength, $\lambda_c$ that corresponds to \Ha\ at $0.02<z<0.2$ and a filter width, $\Delta\lambda$ ranging from $100\Angs$ to $400\Angs$. The design for the first Merian medium-band filter ($N708$), targeting \Ha\ emission, was selected to be centered on $\lambda_c=7080\Angs$ with $\Delta\lambda=275\Angs$, corresponding to $0.057<z<0.103$, following optimization for the lensing S/N and the number of dwarfs in the final sample (see Figure 6 in \citealt{Luo:2023}). The second Merian medium-band filter ($N540$) was matched to the $N708$ filter design, probing \Oiii\ emission within the same redshift range but also avoiding a strong sky emission line at $5580\Angs$. The $N540$ filter is centered on $\lambda_c=5400\Angs$ with $\Delta\lambda=210\Angs$.

The two Merian medium-band filters, $N708$ and $N540$, were fabricated by Asahi Spectra Ltd\footnote{\url{https://www.asahi-spectra.com/}} in 2020-2021 and they have a size of 620 mm in diameter and 14 mm in thickness. The central wavelengths of both filters were measured by Asahi at 49 different locations with 0.5 nm resolution to characterize their uniformity. For $N708$ and $N540$, the uniformity of the central wavelength exhibited a peak-to-valley (p-v) variation of $1.9\,\mathrm{nm}$ (0.27\% of the nominal central wavelength) and $1.9\,\mathrm{nm}$ (0.2\% of the nominal central wavelength), respectively. The FWHMs at those 49 locations were measured to be $27.7\,\mathrm{nm} \pm 0.6\,\mathrm{nm}$ for $N708$ and $21.1\,\mathrm{nm} \pm 0.1\,\mathrm{nm}$ for $N540$. The throughput curves of the two medium-band filters are shown in Figure \ref{fig:trans_curves}, along with the curves for the HSC broad-band filters. These $N708$ and $N540$ throughput curves were generated by averaging the transmission curves measured by Asahi at the 49 different locations on the filters, and by multiplying them with the CCD quantum efficiency, telescope M1 and corrector response curve, and theoretical atmospheric throughput models. 


\subsection{Survey Fields}

\begin{figure*}[t]
 \centering
 \includegraphics[width=1.0\textwidth]{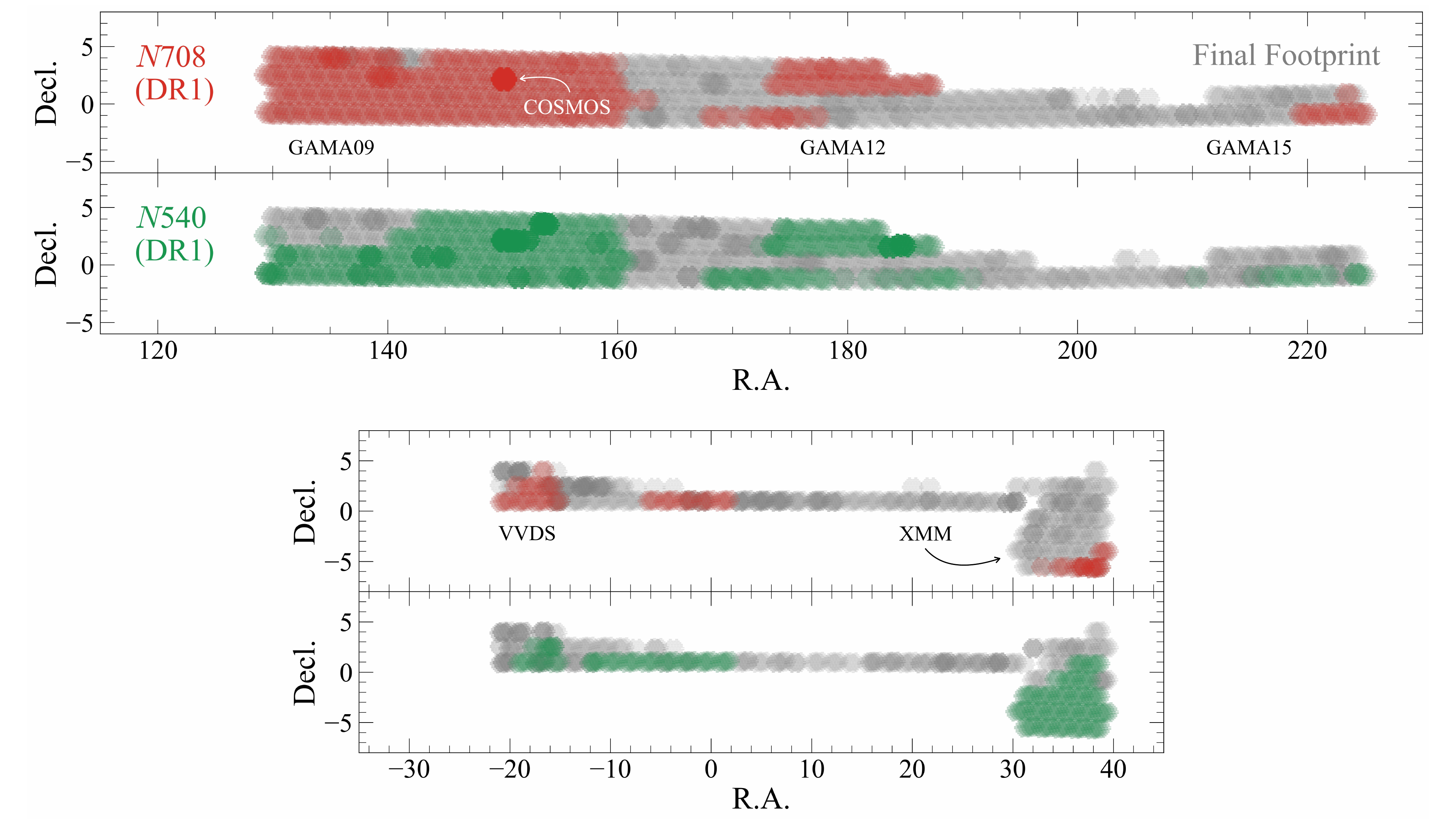}
  \caption{Spring (top panels) and Fall (bottom panels) pointings observed by the Merian Survey. Red and green regions correspond to Merian-Wide pointings observed with the $N708$ and $N540$ medium-band filters, respectively, which are included in DR1. The gray regions outline the footprint of the final survey coverage.}
  \label{fig:survey_map}
\end{figure*}

The Merian Survey consists of a wide layer and a deep layer (Merian-Wide and Merian-Deep, respectively). Merian-Wide targets celestial equatorial fields that were part of the HSC-SSP Wide layer, to enhance scientific synergy with the HSC-SSP broad-band imaging and other publicly available spectroscopic surveys (e.g., GAMA and SDSS). It includes two large, contiguous regions with a combined area of approximately $750\,\mathrm{deg}^2$. The top panel of Figure \ref{fig:survey_map} shows the final survey footprint in gray, including the fall equatorial fields that overlap with the GAMA \citep{Driver:2011} and COSMOS \citep{Scoville:2007} fields, and the spring equatorial fields that overlap with the VVDS \citep{LeFevre:2013}, XMM-LSS \citep{Pierre:2016}, and SXDS fields \citep{Furusawa:2008}.

Merian-Deep covers a single pointing ($\sim 2\,\mathrm{deg}^2$) in the extragalactic COSMOS field, centered at $\alpha=10^\mathrm{h}00^\mathrm{m}28\overset{\mathrm{s}}{.}6$, $\delta=+02^\circ12^\mathrm{m}21^\mathrm{s}$ (J2000). With a total observing time approximately $\times 10$ longer than the nominal Merian-Wide total observing time, Merian-Deep is used for characterizing the selection function and testing photometric measurements and shape systematics within Merian-Wide. Furthermore, it stands as a separate high-quality dataset complementing the extensive array of publicly available deep, multi-wavelength imaging and spectroscopic observations within the COSMOS field \citep[e.g.,][]{Laigle:2016, Weaver:2022}. 

\section{Observations}\label{sec:obs}

\subsection{Observing Strategy}

\begin{figure*}[t]
 \centering
 \includegraphics[width=0.9\textwidth]{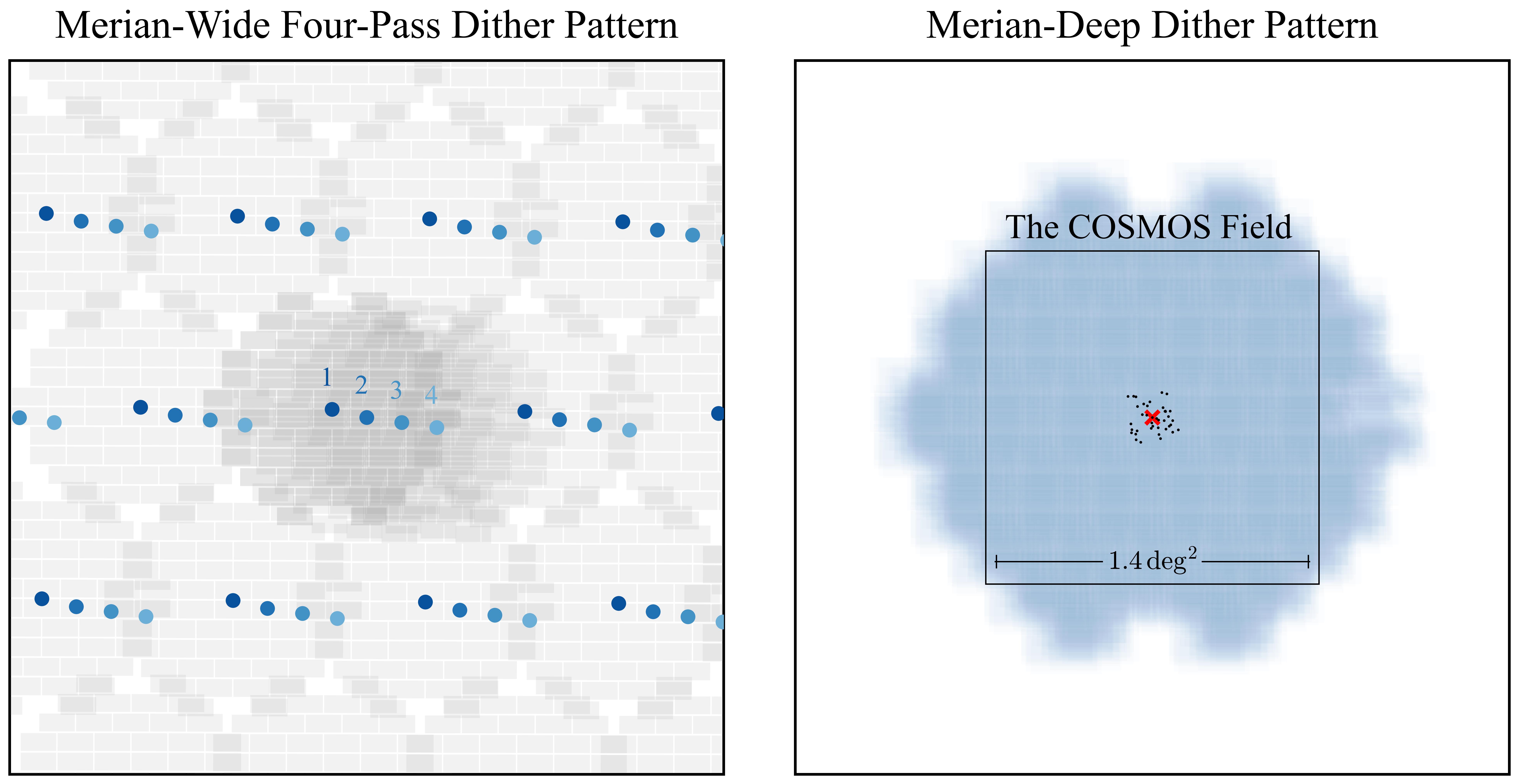}
  \caption{Dither pattern for the Merian survey. In Merian-Wide (left), the survey footprint is covered with four-pass tilings with each tiling pattern offset from the previous one. The left panel shows a region of sky ($6 \times 6\,\mathrm{deg}^2$) covered with the Pass 1 tiling and a full four-pass coverage for a single pointing at the center, demonstrating the full depth. This pattern covers at least $98\%$ of the pointings with four exposures. In Merian-Deep (right), the COSMOS field is observed with $41$ dithered exposures offset from a fiducial pointing (red X mark) centered at $\alpha=10^\mathrm{h}00^\mathrm{m}28\overset{\mathrm{s}}{.}6$ by a radial angular separation of 4.8\arcmin{} and evenly spaced azimuthal offset and an additional random offset in both right ascension and declination drawn from a uniform distribution.}
  \label{fig:dither}
\end{figure*}

The Merian observations in the two optical medium bands ($N708$ and $N540$) were carried out with the DECam on Blanco \citep{Flaugher:2015}. DECam has a mosaicked focal plane comprised of 62 CCDs for imaging, with a total $3.18\,\mathrm{deg}^2$ field of view. As such, wide-field imaging with DECam must consider the gaps between individual CCDs when designing a survey tiling and dither pattern. For Merian-Wide pointing layouts, we have adopted a tiling pattern similar to that used by the DECam Legacy Survey (DECaLS; \citealt{Burleigh:2020}), which employs the precomputed icosahedral arrangements of \citet{Hardin:2012}, uniformly covering a sphere with $N_\mathrm{tiles}=15,872$. We have used a four-pass strategy, consisting of four independent tilings, each offset from the others by a fixed amount. Each of the three subsequent passes employs the same tiling configuration as the initial pass, with each pass shifted by a fixed amount in both right ascension ($\Delta\mathrm{RA}$) and declination ($\Delta\mathrm{Dec}$). Specifically, these offsets are $[-0.2917, 0.0833]\,\mathrm{deg}$, $[-0.5861, 0.1333]\,\mathrm{deg}$, and $[-0.8805,0.1833]\,\mathrm{deg}$, respectively (DECaLS used a similar basic tiling strategy and offsets with three passes compared to the four passes we implement here). Our tiling strategy for Merian-Wide is shown in the left panel of Figure \ref{fig:dither}.

For Merian-Deep, we adopted a modified version of the HSC-SSP deep and ultra-deep dither pattern \citep{Aihara:2018}. We chose a fiducial pointing centered on the COSMOS field. All subsequent pointings were offset from this fiducial pointing by a radial angular separation of 4.8\arcmin{} and an evenly spaced azimuthal offset such that the pointings are distributed uniformly azimuthally. To avoid any persistent chip gap alignments with this strategy, we then applied an additional random offset in both right ascension and declination drawn from a uniform distribution bounded at $\Delta x = \mathcal{U}[-7.5\arcmin,7.5\arcmin]$, where $\Delta x$ is the offset in RA and Dec. The right panel of Figure \ref{fig:dither} shows the dithering pattern of the Merian-Deep layer where 41 exposures are taken with each of the $N708$ and $N540$ filters.

\subsection{Observations}

The Merian survey was originally awarded 62 nights of Blanco/DECam observations with the two custom-made $N708$ and $N540$ filters. To date, \comment{84 nights} have been allocated to the project, compensating for time lost owing to adverse weather conditions or instrumentation issues. \comment{We obtained the first Merian exposures in March 2021, and observations concluded in August 2024.} In Figure \ref{fig:survey_map}, we show the final pointings in gray and the pointings included in the first data release (data release 1; DR1) in red ($N708$) and green ($N540$). The observations included in the survey's first data release DR1 are also summarized in Table \ref{table:observations}. All primary survey fields (Merian-Wide) were observed using both filters, with exposure times of $600\,\mathrm{sec}$ for $N708$ and $900\,\mathrm{sec}$ for $N540$.

We employ the ``effective exposure time'' notion, which unfolded as part of the Dark Energy Survey (DES; \citealt{DES:2016}), to determine the effective depth of each exposure. For every exposure, we calculate the effective exposure time ratio, $\tau$, defined as\footnote{\url{https://lss.fnal.gov/archive/test-tm/2000/fermilab-tm-2610-ae-cd.pdf}}:

\begin{equation}
    \tau = \eta^2 \left(\frac{\mathrm{FWHM}}{\mathrm{FWHM}_\mathrm{canonical}}\right)^{-2} \left(\frac{b}{b_\mathrm{dark}}\right)^{-1},
\end{equation}
where $\eta$ is the atmospheric transmission (with a canonical value of $\eta=1$), $\mathrm{FWHM}$ is the seeing measured as a point spread function (PSF) full width at half maximum (with a canonical value of $\mathrm{FWHM}_\mathrm{canonical}=1''$), $b$ is the measured sky brightness, and $b_\mathrm{dark}$ is the sky brightness representative of zenith dark sky. This factor $\tau$ is then used to calculate an effective exposure time, $t_\mathrm{eff}$, by scaling the open shutter exposure time, $t_\mathrm{exp}$, such that $t_\mathrm{eff}=\tau t_\mathrm{exp}$. We use $\mathrm{FWHM}_\mathrm{canonical}$ of $1''$ and canonical sky brightness values of $21.0\,\mathrm{mag\,arcsec^{-2}}$ and $22.1\,\mathrm{mag\,arcsec^{-2}}$ for the $N708$ and $N540$ filters, respectively. We use the \texttt{copilot} software \citep{Burleigh:2020} to measure the seeing, transparency, and sky brightness. The effective exposure time $t_\mathrm{eff}$ is utilized twice during the observations. Initially, it is used in real-time to decide whether to proceed with the primary Merian program (requiring $t_\mathrm{eff}>200\,\mathrm{sec}$ for $N708$ and $t_\mathrm{eff}>300\,\mathrm{sec}$ for $N540$) or switch to the backup program (see \S \ref{sec:backup}). Additionally, $t_\mathrm{eff}$ values are assessed after the fact to determine if pointing is complete or if it needs to be revisited in a subsequent run. 

\begin{figure*}[t]
 \centering
 \includegraphics[width=0.8\textwidth]{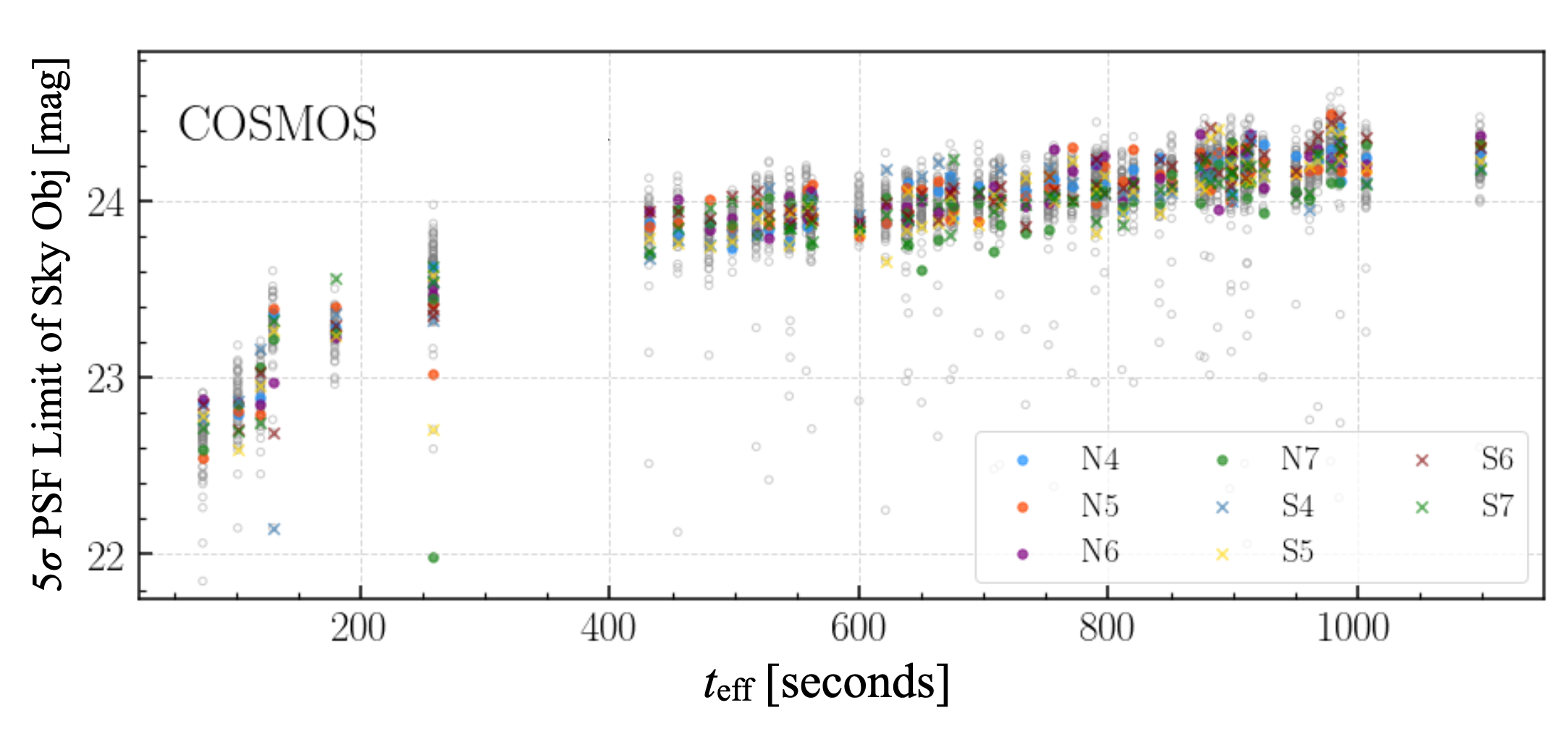}
  \caption{Point source depth for single $N708$ exposures taken in the COSMOS field as a function of their computed $t_\mathrm{eff}$. Each point corresponds to the depth measured from one of the DECam CCDs, e.g., N4, S7.}
  \label{fig:detect_lim}
\end{figure*}

\begin{figure*}[t]
\centering
\begin{subfigure}
    \centering
    \includegraphics[width=1.0\textwidth]{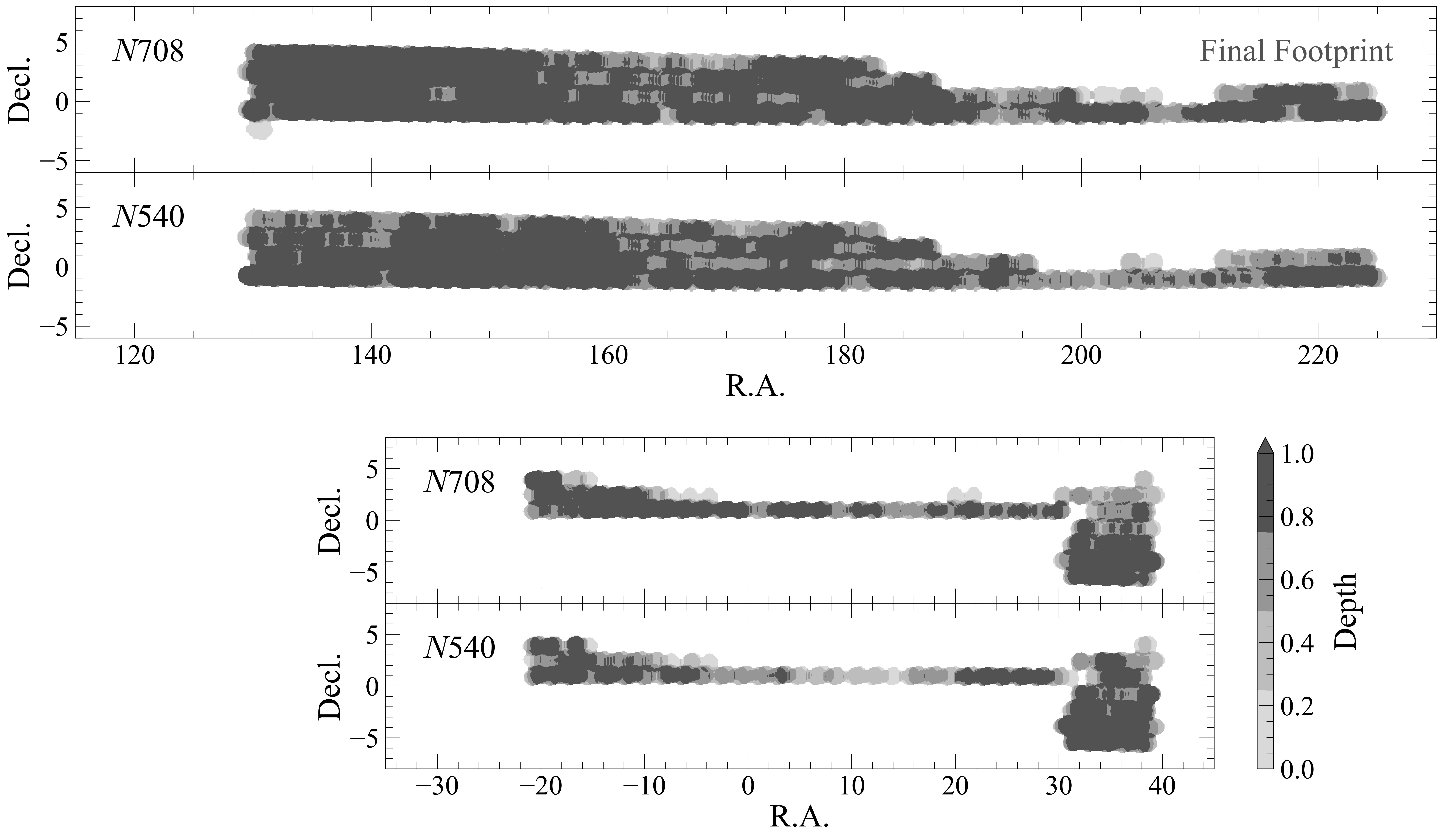}
    \end{subfigure}
 
 \vspace{1em} 
 
\begin{subfigure}
     \centering
     \includegraphics[width=1.0\textwidth]{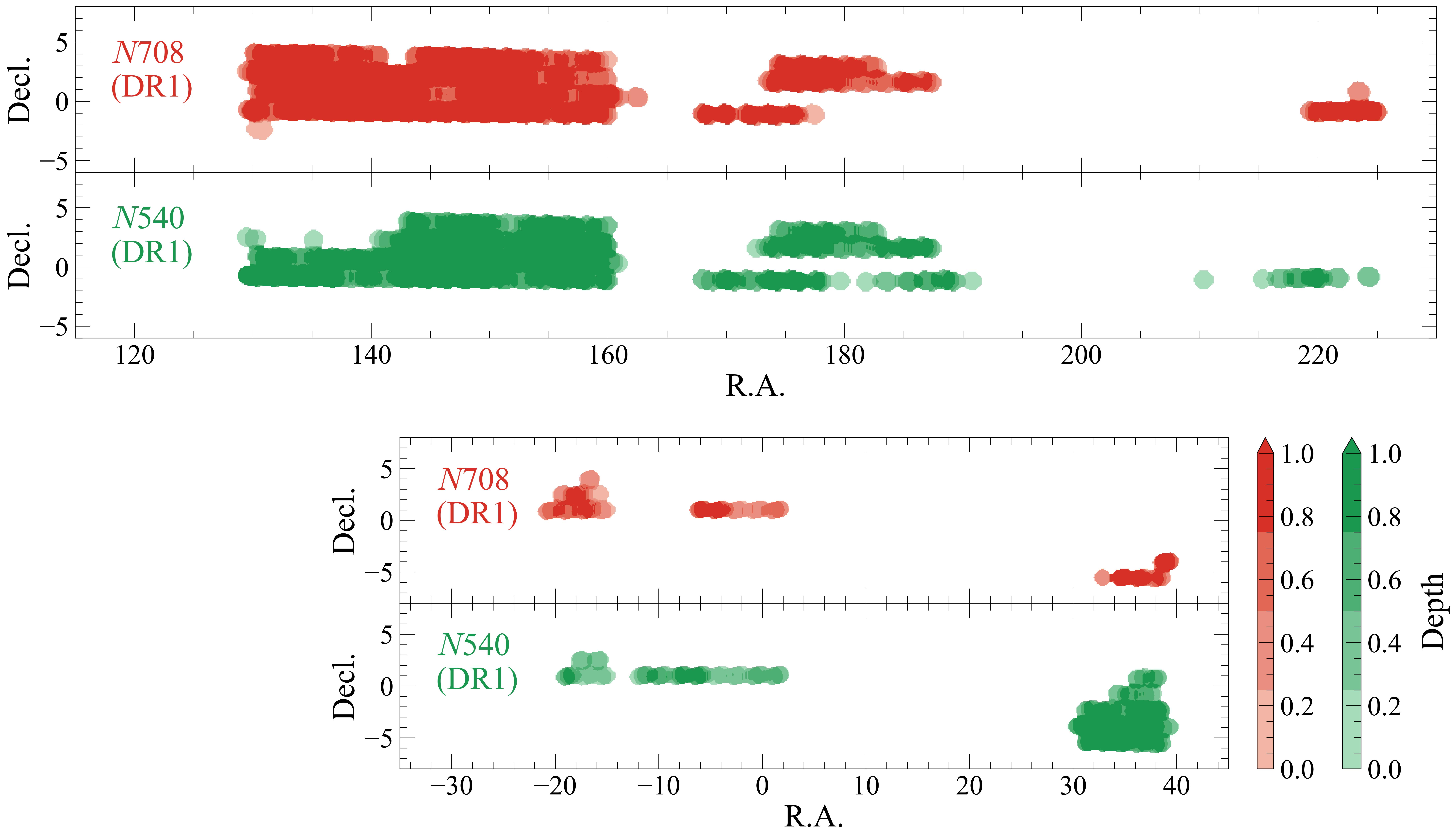} 
\end{subfigure}

\caption{\comment{The Merian Survey effective exposure time ($t_\mathrm{eff}$) outlining the Spring and Fall footprints for the $N708$ and $N540$ filters. The upper panels show the final footprint in gray shades whereas the bottom panels show the DR1 footprints. In all panels, darker colors (gray, red, and green) represent the deepest data where Depth=1 corresponds to $t_\mathrm{eff}\geq2400\,\mathrm{sec}$ for $N708$ and $t_\mathrm{eff}\geq3600\,\mathrm{sec}$ for $N540$.}}
\label{fig:t_eff_dist}
\end{figure*}

\begin{figure*}[t]
 \centering
 \includegraphics[width=0.9\textwidth]{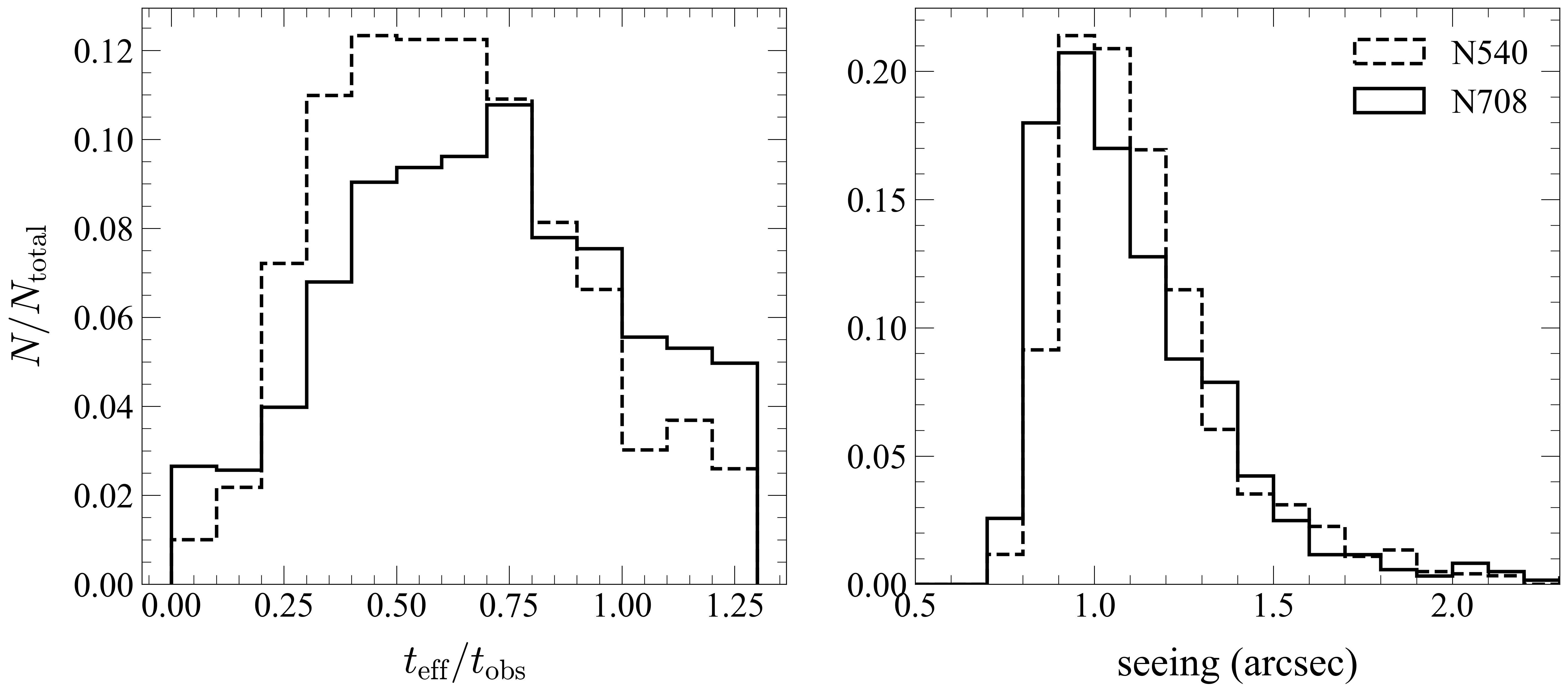}
  \caption{Effective exposure time ($t_\mathrm{eff}$) normalized by nominal exposure times (left; $t_\mathrm{exp}=900\,\mathrm{sec}$ for $N540$ and $600\,\mathrm{sec}$ for $N708$) and seeing distribution (right) for the DECam $N708$ and $N540$ exposures included in Merian-Wide DR1.}
  \label{fig:t_eff_seeing_dist}
\end{figure*}

In Figure \ref{fig:detect_lim}, we show the $5\sigma$ point source depth for single $N708$ exposures as a function of their computed $t_\mathrm{eff}$. $t_\mathrm{eff}$ spans a wide range of values between $\sim 50$ to $1000\,\mathrm{sec}$ for a fixed exposure time of $t_\mathrm{exp}=600\,\mathrm{sec}$. As expected, the point source depth increases as a function of increasing $t_\mathrm{eff}$. Single $N708$ exposures with $t_\mathrm{eff}<200\,\mathrm{sec}$ are retaken. \comment{Figure \ref{fig:t_eff_dist} shows the $t_\mathrm{eff}$ maps for all fields in both Merian filters, displaying the final survey footprint in the upper panels and the DR1 coverage in the lower panels.} Darker regions represent the deepest data (highest $t_\mathrm{eff}$ with Depth=1 corresponding to $t_\mathrm{eff}>2400\,\mathrm{sec}$ for $N708$ and $t_\mathrm{eff}>3600\,\mathrm{sec}$ for $N540$). Figure \ref{fig:t_eff_seeing_dist} presents the distribution of $t_\mathrm{eff}$ normalized by the nominal exposure time for the two filters (left) and the PSF FWHM distribution (right), all based on the Merian-Wide DR1 observations. With $t_\mathrm{eff}$ in hand, we generate a Hierarchical Equal Area isoLatitude Pixelization (\texttt{HEALPix}; \citealt{Gorski:2005}) projection map for DR1. The \texttt{HEALPix} mask includes regions with data in all seven bands ($grizyN540N708$; full color) and with $t_\mathrm{eff}>1200\,\mathrm{sec}$ for $N708$ and $>1800\,\mathrm{sec}$ for $N540$, defined as full depth. \comment{The full-color (FC) area of the survey is $\sim 733\,\mathrm{deg}^2$ with a full color-full depth (FCFD) area of $\sim 584\,\mathrm{deg}^2$. DR1 has an FC area of $320\,\mathrm{deg}^2$ and an FCFD of $234\,\mathrm{deg}^2$.}

\begin{table*}[htbp]
\setlength{\tabcolsep}{9pt}
\renewcommand{\arraystretch}{1.2}
\caption{Observations to Date}
\centering 
\begin{tabular}{c c c c} \hline\hline 
\multicolumn{4}{c}{\textbf{Data Release 1}} \\
Observing Block Dates & $\#$ of Nights Observed & Filters & Fields \\
\hline 
March 2021 & 6 & N708 & COSMOS, GAMA \\
\hline 
September 2021 -- January 2022 & 15 & N708, N540 & XXM, VVDS, SXDS \\
\hline 
February 2022 -- March 2022 & 12 & N708, N540 & GAMA \\
\hline 
& & & \\ 
\multicolumn{4}{c}{\textbf{Post DR1 Observations}} \\
Observing Block Dates & $\#$ of Nights Observed & Filters & Fields \\
\hline 
September 2022 -- October 2022 & 6.5 & N708, N540 & XXM, VVDS, SXDS \\
\hline 
March 2023 -- May 2023 & 18 & N708, N540 & GAMA \\
\hline 
August 2023 -- November 2023 & 17 & N708, N540 & XXM, VVDS, SXDS \\
\hline 
April 2024 & 6.5 & N708, N540 & GAMA \\
\hline 
August 2024 & 3 & N708, N540 & XXM, VVDS \\
\hline 
\hline 
\end{tabular}
\label{table:observations}
\end{table*}

\subsection{Spectroscopic Observations}\label{sec:merian_spec}

In addition to the main imaging campaign, there is also an ongoing spectroscopic effort to calibrate and validate the accuracy of the seven-band photometric redshifts. We first compiled and vetted existing spectroscopic data from two publicly available surveys, SDSS and GAMA. We additionally acquired new spectroscopic data for potential in-band low-mass galaxies in the extragalactic COSMOS field. We selected galaxies with $7.5<\log(M_\star/M_\odot)<9.5$ and $z<0.25$ from the COSMOS2015 catalog \citep{Laigle:2016} as our primary spectroscopic targets. We also included a sample of \Lya\ emitter candidates at $z\sim3.4$ and $z\sim 4.8$ selected via their $N540$ and $N708$ flux excess, respectively.

We obtained spectroscopic data with the wide-field multislit Inamori-Magellan Areal Camera and Spectrograph \citep[IMACS;][]{Dressler:2006} on the Magellan Baade Telescope through a sequence of observational programs (PIs: T.Li \& S.Danieli) from January 2020 to February 2022. We used the $f/2$ camera on IMACS to maximize the field of view and the 200 line/mm grism combined with the GG495 blocking filter to cover the wavelengths from 5000 to 9000\AA. Overall, we observed 8 slitmasks with 1.0 arcsec slit width and $\sim$ 3 hours of exposure per mask. We also obtained data with the Deep Extragalactic Imaging Multi-Object Spectrograph \citep[DEIMOS;][]{Faber:2003} on the Keck-II telescope between February 2022 and March 2023 (PI: A.Leauthaud, PI: E.Kado-Fong). Two different gratings were used for the Keck/DEIMOS programs. The 600ZD grating provides a wide wavelength coverage from 4500 to 9600\AA\ and a spectral resolution of R$\sim$2000. The 1200G grating covers the wavelengths from 5700 to 8300\AA, with a higher spectral resolution (R$\sim$4000). We observed 13 slitmasks with the 1200G grating and 12 slitmasks with the 600ZD grating with 1.0 arcsec slit width and 1 hour of exposure per mask. In addition to the Magellan/IMACS and Keck/DEIMOS observations, we collected spectroscopic data in the COSMOS field through an ancillary program with the Dark Energy Spectroscopic Instrument \citep[DESI;][]{DESI:2022} on the Mayall Telescope at Kitt Peak National Observatory in March 2023. DESI is a robotic and fiber-fed spectroscopic instrument with a wide field of view. It covers a wide wavelength range from 3600 to 9800\AA, with a high efficiency. We broke the program into 22 observations with a nominal exposure time of 1000s each with a dithered tile to limit the effects of instrumental artifacts. The average exposure time of the DESI spectra is $\sim$2.2 hours.

In total, we have collected spectra for 3,914 dwarf galaxies ($7.5<\log(M_\star/M_\odot)<9.5$) in the COSMOS field using the IMACS, DEIMOS, and DESI observations described above. These spectra are used to characterize the photometric redshifts derived using the Merian seven-band photometry. Additional information regarding the Merian photometric redshifts and spectroscopic data will be provided in Luo et al. in preparation.

\subsection{The Merian Backup Program}\label{sec:backup}

As mentioned in Section~\ref{sec:obs}, when the effective exposure times dropped below our thresholds, we conducted observations for the Merian backup program. There are two components to the backup program: extending time domain observations taken by HSC and monitoring known low-mass active galactic nuclei (AGN). The first aspect of the backup program aims to extend the baseline of time domain data taken by HSC in the COSMOS and SXDS fields. Observations were conducted in the $g$ and $r$ bands with exposure times set at either $90\,\mathrm{sec}$ or $300\,\mathrm{sec}$ for both filters. Priority was given to $g$ band observations. A total of 22 pointings were observed in the COSMOS field and eight in the SXDS field.

The second aspect of the backup program entailed monitoring a sample of known low-mass AGN for short-timescale variability. We selected targets exhibiting optical spectroscopic AGN signatures suggestive of a low-mass black hole ($M_{\rm BH}\approx 10^{5-7} M_{\odot}$). Depending on the specific target, we conducted exposures of either 90 seconds or 300 seconds in the $g$ band. Typically, each target was observed for approximately two hours in total on a given night. We made between three and five visits to each target in total.

\section{Data Reduction}\label{sec:data-reduction}

Merian uses two sets of optical imaging data: new medium-band data ($N708$, $N540$) obtained with the Dark Energy Camera on the Blanco telescope, as described in Section~\ref{sec:survey_design}, and the five broad-band ($grizy$) data in the HSC-SSP Public Data Release 3 (PDR3, also known as S20A; \citealt{Aihara:2022})\footnote{\url{https://hsc-release.mtk.nao.ac.jp/doc/index.php/available-data__pdr3/}}. We opted to perform our own data reduction for the newly acquired DECam data instead of relying on data processed using the DECam community pipeline\footnote{\url{https://noirlab.edu/science/index.php/data-services/data-reduction-software/csdc-mso-pipelines/pl206}}. This choice provides the benefit of guaranteeing improved compatibility with the five broad-band HSC data, which is processed using the LSST Science Pipelines (see below), and allows us to perform forced, aperture-matched photometry across all seven bands, as further detailed below (\S \ref{sec:photometry-catalog}).

The HSC and DECam data sets were reduced using different adaptations of the Rubin Observatory LSST Science Pipelines\footnote{\url{https://pipelines.lsst.io/}}, tailored to support the Subaru Telescope’s HSC instrument and the Blanco's DECam instrument data reduction, respectively. The flexibility and extensibility inherent in the LSST Science Pipelines software architecture enable its adaptation for reducing data acquired with the DECam. While the HSC-SSP data were processed as part of PDR3 \citep{Aihara:2022}, we conducted the DECam data reduction using a customized version of the LSST Science Pipelines. This is the first attempt to reduce an extensive DECam dataset using the LSST Science Pipelines.

The most thorough description of the pipelines, as developed for processing data from the HSC instrument, is given in \citet{Bosch:2018, Bosch:2019} and \citet{Jenness2022}. Here we describe the high-level image processing procedures of the LSST Science Pipelines and describe the choices and customizations that were made during the processing of the DECam data.

\subsection{Merian DECam Data Reduction using the LSST Science Pipelines}\label{sec:lsst-scipipe}

All of the data reduction tasks described below were performed using the \texttt{w\_2022\_29} tagged version of the LSST Science Pipelines (i.e., the snapshot of the Science Pipelines released on week 29 of 2022)\footnote{\url{https://pipelines.lsst.io/v/w_2022_29/index.html}}. A more detailed description of LSST Science Pipelines can be found in \citet{Bosch:2019}.

\vspace{0.2cm}
\noindent
\textbf{Single frame processing (CCD Processing).} Individual raw frames are processed through Instrument Signature Removal (ISR), including their flat-fielding, bias subtraction, fringe correction, nonlinearity and crosstalk correction, and masking of bad and saturated pixels. Next, single-epoch direct image characterization is done, including sky background subtraction based on an initial model of the sky background, PSF modeling from bright stars, detecting and interpolating over cosmic rays, applying aperture corrections, deblending, and creating a source catalog for individual frames. We note that many of these steps are run semi-iteratively as described in \citet{Bosch:2018}. Sources in the source catalog are then used for performing astrometric and photometric calibration where the Gaia DR2 catalog \citep{Gaia:2016} is used for astrometric calibration and Pan-STARRS PS1 \citep{Chambers:2016} for photometric calibration. 

\vspace{0.2cm}
\noindent
\textbf{Global Calibration.} Once individual exposures have undergone calibration and characterization, steps are applied during joint processing to enhance the overall reduction outcome. In particular, the sky background modeling and the steps handling artifacts (cosmic rays, satellite trails, and optical ghosts) are repeated through an image differencing analysis, utilizing the dithering between individual exposures. Next, the astrometric and photometric calibrations are improved by: (1) using a larger number of calibration sources; and (2) requiring a solution when positions and fluxes are measured in different locations of the focal plane and during different visits. 

\begin{figure*}[t]
 \centering
 \includegraphics[width=1.0\textwidth]{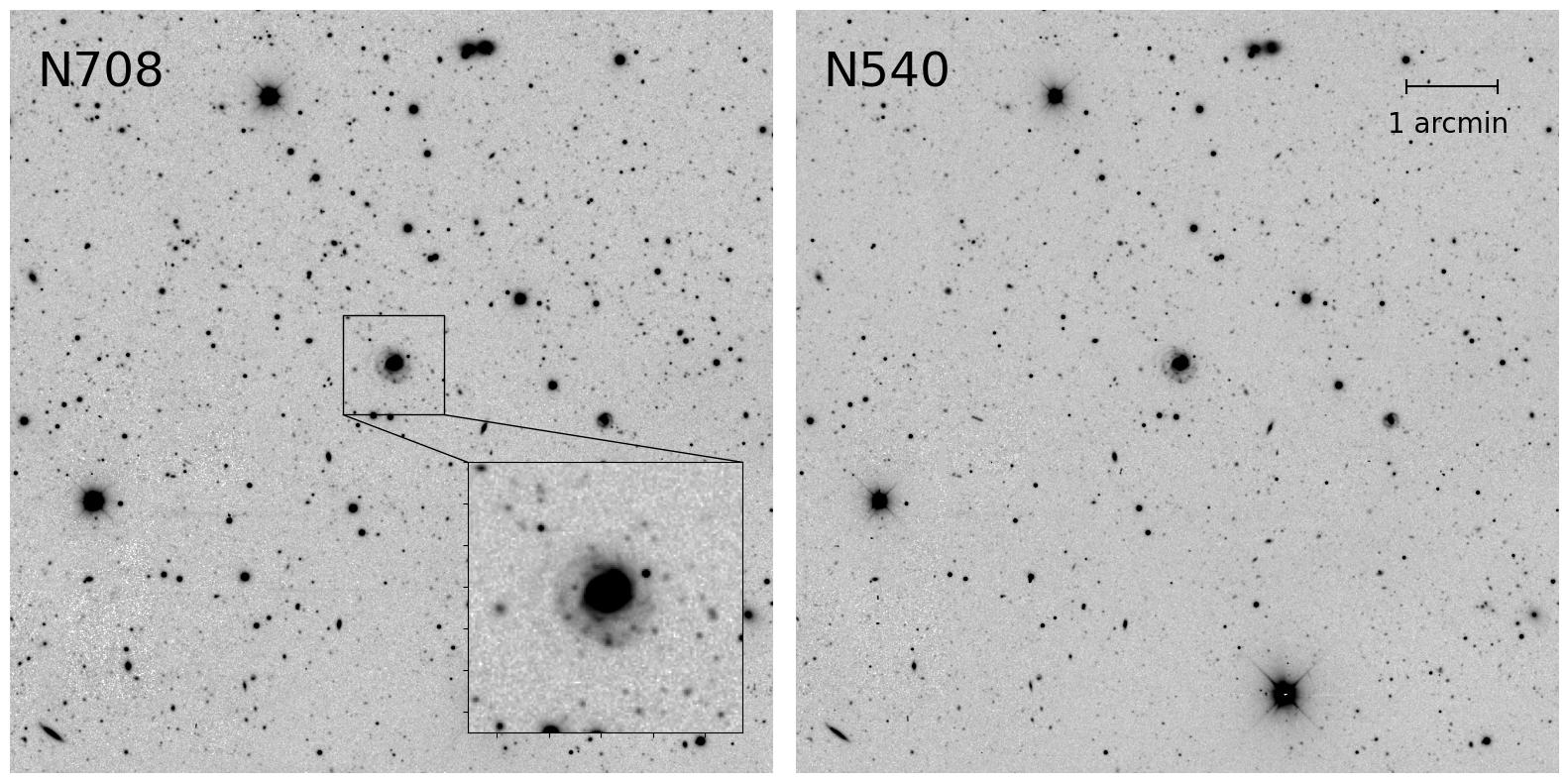}
  \caption{Example $N708$-band (left) and $N540$-band (right) coadd cutouts ($\sim 8.3 \times 8.3\,\mathrm{arcmin}$) from the Merian DR1 dataset, shown using a square root stretch. Both cutouts are centered at $\alpha=10^\mathrm{h}11^\mathrm{m}57\overset{\mathrm{s}}{.}58$, $\delta=+00^\circ58^\mathrm{m}58.44^\mathrm{s}$ (J2000)}
  \label{fig:example_img}
\end{figure*}

\vspace{0.2cm}
\noindent
\textbf{Image coaddition.} Coadded images (`coadds') are constructed by a direct weighted average of resampled individual frames to a common pixel grid. For this purpose, we opt to resample Merian DECam data onto the skymap adopted by HSC. The HSC rings skymap divides the sky (or a wide region of the sky) into tracts of $1.68 \times 1.68\,\mathrm{deg}^2$ which are further subdivided into $9 \times 9$ patches which are each $\sim 12\,\mathrm{arcmin}$ on the side. Coadds for each patch and filter are constructed independently. As a consequence of remapping onto an HSC-derived skymap, DECam data is oversampled by a factor of $\sim 1.6$ compared to the DECam native pixel size of $0.263\,\mathrm{arcsec/pixel}$\footnote{We also reduced several tracts using DECam skymaps with the DECam native pixel scale and found no significant difference in terms of photometric (and astrometric) precision.}. In Figure \ref{fig:example_img}, we show examples for $N708$ and $N540$ coadd images.

\subsection{Image and Reduction Quality Assurance}

\subsubsection{Astrometric and Photometric Calibrations}

\begin{figure*}[t]
 \centering
 \includegraphics[width=1.0\textwidth]{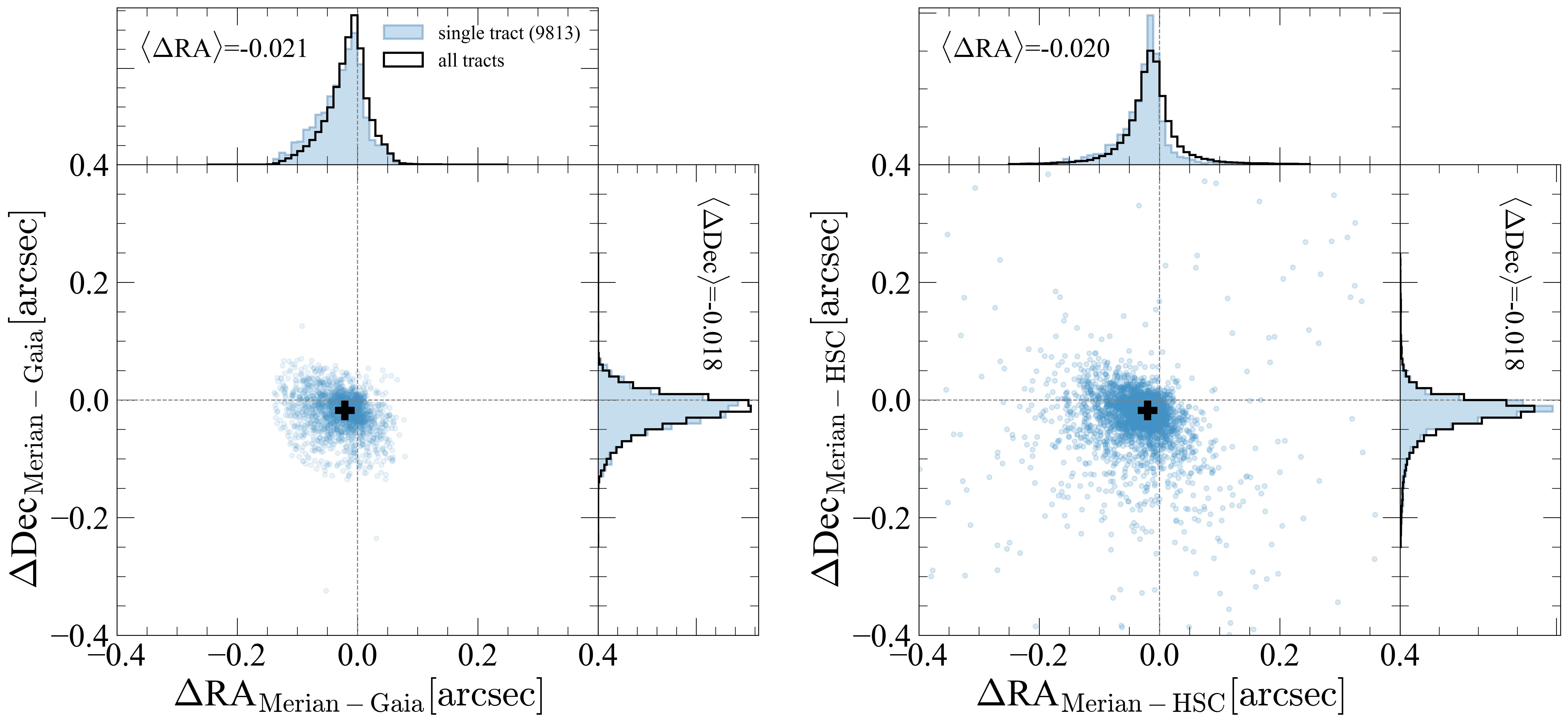}
  \caption{Astrometric offset in right ascension and declination for a representative tract (light blue) and a larger portion (10\%) of the survey DR1 data (black histogram). We match stars between Merian and Gaia (left) and HSC (right) and compute the offset.}
  \label{fig:astro_calib}
\end{figure*}

We test the astrometric calibration by comparing the Merian, HSC, and Gaia DR3 coordinates of common stars with $m_g<20\,\mathrm{mag}$ based on the CModel photometry. In Figure \ref{fig:astro_calib}, we show the computed offset in R.A. ($\Delta\mathrm{RA}$) and Dec. ($\Delta\mathrm{Dec}$) when comparing to the stars in Gaia-DR3 (left) and in HSC-SSP (right). The light blue points and histogram show stars in a single tract and the black histogram shows the median offset in R.A. and Dec. for stars from $10\%$ of the tracts in Merian-DR1. Overall, the astrometric calibrations are good with median values of $\langle\Delta(\mathrm{RA})\rangle=-0.020\,\mathrm{arcsec}$ and $\langle\Delta(\mathrm{Dec})\rangle=-0.018\,\mathrm{arcsec}$. 



\begin{figure*}[t]
 \centering
 \includegraphics[width=1.0\textwidth]{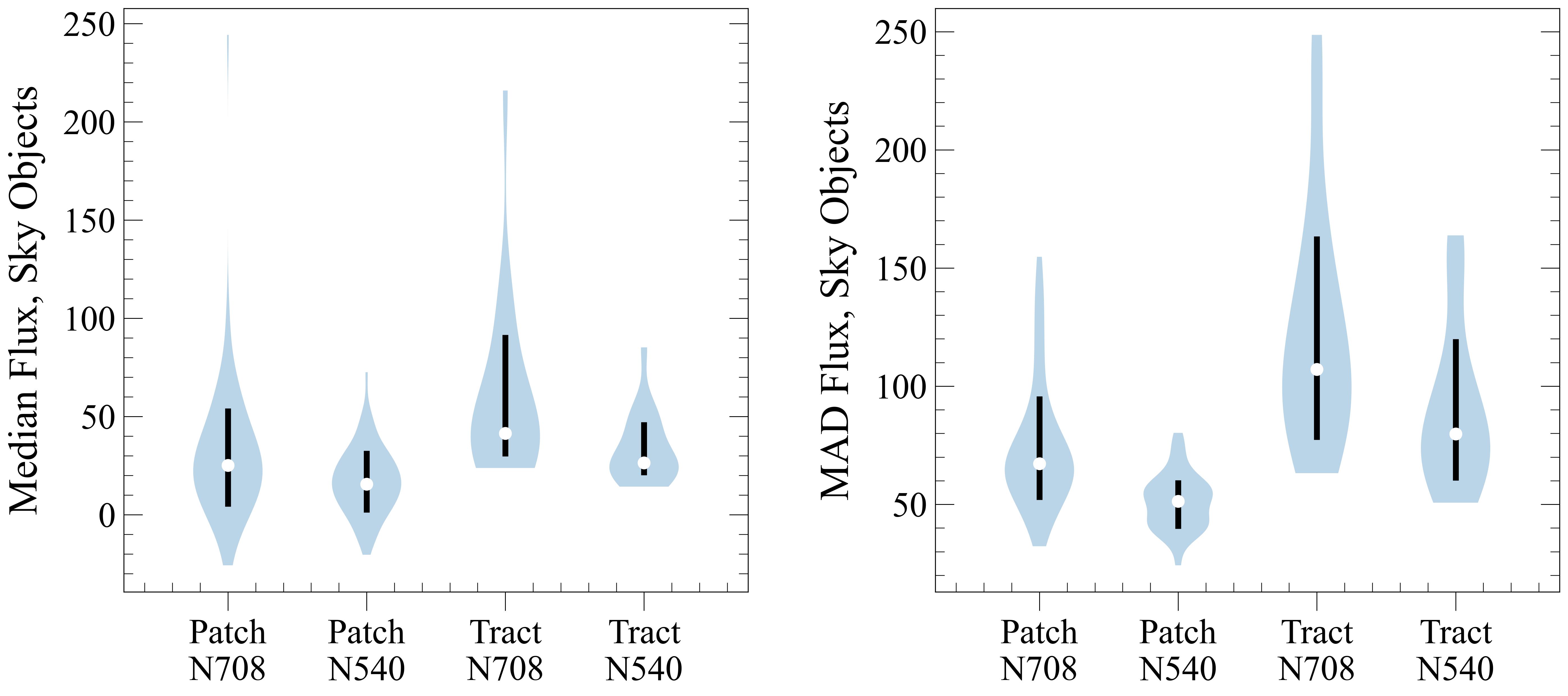}
  \caption{The flux distribution of sky objects in the $N540$ and $N708$ bands. The \texttt{N708\_gaap1p0Flux} and \texttt{N540\_gaap1p0Flux} fluxes are used. We show the patch-level and tract-level median (left) for sky objects in a representative tract (tract 8525) and a larger set of tracts, respectively, along with their corresponding median absolute deviation (MAD, right). The black bars represent the range from the 16th to 84th percentiles, and the white dots indicate the 50th percentile. These test results indicate good sky subtraction quality, with flux values for the sky objects remaining close to zero in both bands.}
  \label{fig:skyobj_test}
\end{figure*}


\subsubsection{Sky Background Subtraction}
The LSST Science Pipelines utilizes `Sky Objects' for testing the sky background in single exposures and coadded images. Sky Objects are small empty regions in which no real objects are detected. These regions are selected to be free of any significant light sources so that the measured flux can represent the true background level. Fluxes with different aperture sizes are measured for these artificial ``objects'' and added to the coadd level object table. These objects are eventually removed from the final photometric catalog (see Section \ref{sec:photo_cat}), however, they are used as indicators of the local sky background. Figure \ref{fig:skyobj_test} shows median (left) and mean absolute deviation (right) $N708$ and $N540$ fluxes for sky objects in a representative set of tracts from DR1 at the patch and tract level. Overall, the quality of sky subtraction looks good with values close to zero.

\subsection{Possible Future Improvements to the Data Reduction}
\comment{The upcoming data reduction phase will occur during the final data release, where all survey data will be processed collectively. We expect to continue using the LSST Science Pipelines, leveraging a newer version that will be released and tested shortly before data processing begins. Reflecting on the DR1 processing, several enhancements can be made. First, each new weekly version of the LSST Science Pipelines builds on its predecessor, integrating the latest algorithmic improvements, bug fixes, and performance optimizations. Potential advancements in calibration may include implementing the Forward Global Calibration Method (FGCM; \citep{Burke:2018}) for photometric calibration, which uses a forward model approach based on atmospheric model parameters and scans of instrument throughput as a function of wavelength. For astrometric calibration, we plan to test the instrumental signature fitting and processing program GBDES \citep{Bernstein:2017}. Additionally, we aim to explore the transition to full-focal-plane sky modeling and background subtraction, and modification of the default deblending parameters to enhance source detection using the \texttt{Scarlet} software \citep{Melchior2018} which we expect will further improve source shredding. The next data release will ensure consistency across both Merian-Deep and Merian-Wide, with any modifications uniformly applied to all datasets in each release.}

\section{Photometry and Photometric Catalogs}\label{sec:photometry-catalog}

We construct the Merian photometric catalog as detailed below, largely following the techniques discussed in depth in \citet{Bosch:2018} and \citet{Bosch:2019}. In summary, we use the $N708$-band coadds for deblending and detection (and the $N540$-band coadds where the $N708$ data are unavailable). Centroids and shape-related measurements are also performed on the $N708$-band images where such imaging exists; if $N708$ imaging is unavailable, the $N540$-band images are used for these measurements. We then perform forced photometry on the rest of the bands using these fixed measurements -- the $N540$-band images from the DECam observations and the $grizy$ from the HSC-SSP observations. As described below, the LSST Science Pipeline performs various multi-band photometric measurements.

\subsection{Source Detection, Deblending, and Measurements}

The procedures for detection and deblending are primarily based on the methods outlined in \citet{Bosch:2018}. Here, we provide a short overview of the key steps and emphasize any difference from \citet{Bosch:2018}. Following coaddition, the detection pipeline runs independently on $N708$ and $N540$ data, wherein objects are identified through a $5\sigma$ threshold applied after Gaussian kernel smoothing with the kernel size matched to the PSF in each band. The detected ``footprints'' in both bands are then merged to eliminate spurious detections. Given that each footprint may encompass multiple peaks corresponding to distinct astrophysical sources, the deblending pipeline is executed to allocate the total flux to each peak. 

Unlike the deblending process described in the HSC pipelines \citep{Bosch:2018, Aihara:2022}, the recent LSST Science Pipelines integrate \texttt{Scarlet} \citep{Melchior2018} as the default deblender\footnote{\url{https://github.com/lsst/scarlet}}. Whereas the deblending algorithm implemented in HSC and SDSS \citep{Lupton2001} operates solely on single-band data, \texttt{Scarlet} leverages color and morphology information to separate blended objects in a non-parametric manner. \texttt{Scarlet} has been demonstrated to outperform single-band approaches in deblending complex scenes. Following the deblending of the ``parent'' footprint, each peak generates its own ``child image'', encompassing both real astrophysical flux and noise. These deblended child images are utilized for the measurement of source properties. During the measurement process, each footprint in the image is substituted with random noise, and the deblended child image corresponding to a particular source is put back when the pipeline computes the properties for that source. This procedure is repeated for all the sources in the image. The basic measurements include the centroids, shapes, PSF photometry, CModel photometry, and various aperture photometry as described below. Blendedness, which measures the contribution of other sources in the neighborhood of a source, is also included.

\subsection{Photometry}

We use the LSST Science Pipelines to perform joint photometry of sources in the two DECam medium-band images and the five HSC broad-band images. The LSST Science Pipelines provides a variety of source photometry including PSF and Kron photometry, CModel photometry (see \citealt{Bosch:2018}), and fixed-aperture photometry. Recently introduced, it also provides measurements of aperture-matched photometry, implementing the ``Gaussian-Aperture-and-PSF'' (GAaP) technique described in \citet{Kuijken:2008}. GAaP performs PSF- and aperture-matched photometry, optimized for measuring accurate galaxy colors from images taken under different PSFs and seeing conditions, and even across multiple telescopes~\citep{Hildebrandt:2020}. This photometry is well-suited to our survey, which includes seven band data from two different photometric systems. Measuring ``simple'' aperture photometry would result in large systematic errors when deriving photometric redshifts as those require accurate color measurements. 

GAaP is described in detail in \citet{Kuijken:2008} and was successfully used in weak lensing surveys such as the Kilo-Degree Survey (KiDS; e.g., \citealt{Hildebrandt:2017, Hildebrandt:2020}). We summarize the fundamental principle and direct the reader to \citet{Kuijken:2008} for an in-depth description of the algorithm and to LSST DMTN-190\footnote{{\url{https://dmtn-190.lsst.io/}}} for the implementation details. In the first measurement step, all detection footprints, except for that of the object being measured, are replaced with noise. The PSF is evaluated at the object's centroid, and a local PSF-matching kernel is derived based on its size. A sufficiently padded subimage around the object is then convolved with this kernel. This convolution results in a Gaussian-shaped PSF for each object, causing a slight degradation in the FWHM. Subsequently, aperture photometry is performed on this PSF-Gaussianized coadd image using a Gaussian aperture weight function. The aperture size is chosen to yield a consistent value for all objects when combined in quadrature with the Gaussian PSF size.

GAaP photometry and other photometric and source measurements are automatically performed on the newly obtained $N708$ and $N540$ images as part of the data processing with the LSST Science Pipelines. Nevertheless, \texttt{Scarlet} deblending and GAaP photometry were not available for the HSC-SSP S20A data we use. While future HSC-SSP data releases will incorporate \texttt{Scarlet} deblending and GAaP photometry, the differing depths and resolutions between DECam and HSC may complicate direct catalog matching. Therefore, we use the DECam footprint for performing the GAaP photometry on the HSC-SSP S20A images. We note that although DECam has lower resolution and shallower depth compared to HSC, which may result in suboptimal detection and deblending when using the DECam footprint for the HSC data, using the same deblended footprints across all seven bands in the ``forced'' mode ensures consistent color measurements. We download HSC-SSP S20A data in the overlapping tracts and run the GAaP measurements taking the deblended footprints from DECam $N708$+$N540$ data. Similarly, we also measure CModel photometry on HSC-SSP S20A data using the DECam footprints. However, we do not independently measure the shapes and blendedness for the HSC-SSP S20A images, as these measurements are more upstream and require running \texttt{Scarlet}.

\subsection{Photometric Catalog}\label{sec:photo_cat}

We construct the Merian photometric catalog as follows. First, we merge the two photometric catalogs, namely the source measurement output from the LSST Science Pipelines for the $N708$ and $N540$ images and the output from the standalone source measurement performed on the HSC-SSP S20A $grizy$ images as described above. To ensure the catalog remains manageable, we opt to keep only a select number of output columns essential for a wide range of scientific applications, data reduction, and photometric quality assurance tasks.\footnote{An example of the complete output catalog columns from the LSST Science Pipelines can be found at \url{https://dm.lsst.org/sdm_schemas/browser/dp02.html}.}

Next, we select unique sources using the \texttt{detect\_isPrimary} flag. Filtering sources according to this flag ensures that a source is deblended (compared to original, blended sources), that the source is located in the interior of a patch and tract rather than in patch overlaps and hence might appear in the catalog more than once, and finally that it is not a sky object. We then apply an SNR cut, retaining only those sources with an SNR greater than 5 in both the $N708$ and $N540$ images, utilizing the \texttt{N708\_gaap1p5Flux} and \texttt{N540\_gaap1p5Flux} fluxes along with their respective uncertainties. Next, we apply the HSC-SSP S20A/21A bright star mask to the catalog, marking sources with unreliable photometry due to saturated stars using the \texttt{IsMask} flag. This step results in approximately 5\% of the sources being flagged as masked.

We include in the photometric catalog measurements from the HSC-SSP S20A catalog that were not performed as part of our independent source measurement process on the $grizy$ images. These measurements contain parameters like size, blendedness, and extendedness. Specifically, we query all of the unique sources in the HSC-SSP S20A catalog, tract by tract, and match them to sources in our photometric catalog. The matching process involves identifying the nearest on-sky counterparts of an object in the Merian photometric catalog within a set of coordinates from the HSC-SSP S20A catalog, with matched objects required to have an on-sky separation of no more than $1''$. To preserve the origin of each measurement and maintain clarity regarding their respective data sources, we add suffixes to the column names, thereby indicating whether the measurements were derived from our LSST Science Pipeline runs (\texttt{\_Merian} suffix) or obtained from the HSC-SSP S20A catalog (\texttt{\_HSC} suffix).

Finally, we include two flags in the catalog to facilitate the straightforward selection of sources with consistently reliable photometry (\texttt{PhotUse}) and science-ready sources (\texttt{SciUse}). The \texttt{PhotUse} is set to 1 when a source is not masked by the bright star mask. The \texttt{SciUse} is set to 1 when the following criteria are satisfied: 
\begin{enumerate}[itemsep=-0.7cm]
    \item The source is not masked: \\ 
    \vspace{0.5cm}
    \texttt{IsMask\_Merian} = 0
    \vspace{0.5cm}
    \item The source is in the full-color full-depth area. \\ 
    \vspace{0.5cm}
    \item The source meets the following quality criteria: \\ 
    \vspace{0.05cm}
    \texttt{cModel\_flag\_Merian} = 0 \\
    \vspace{0.05cm}
    \texttt{pixelFlags\_edge\_Merian} = 0 \\
    \vspace{0.05cm}
    \texttt{interpolatedCenter\_Merian} = 0 \\
    \vspace{0.05cm}
    \texttt{centroid\_flag\_Merian} = 0 \\
    \vspace{0.05cm}
     \vspace{0.5cm}
    \item Not a star according to its $i$-band extendedness value used for star-galaxy separation\footnote{\url{https://hsc-release.mtk.nao.ac.jp/doc/index.php/star-galaxy-separation__pdr3/}}: \\ 
    \vspace{0.05cm}
    \texttt{i\_extendedness\_value\_HSCS20A} = 1 \\
\end{enumerate}

\vspace{-0.5cm}

\begin{figure*}[t]
 \centering
 \includegraphics[width=0.6\textwidth]{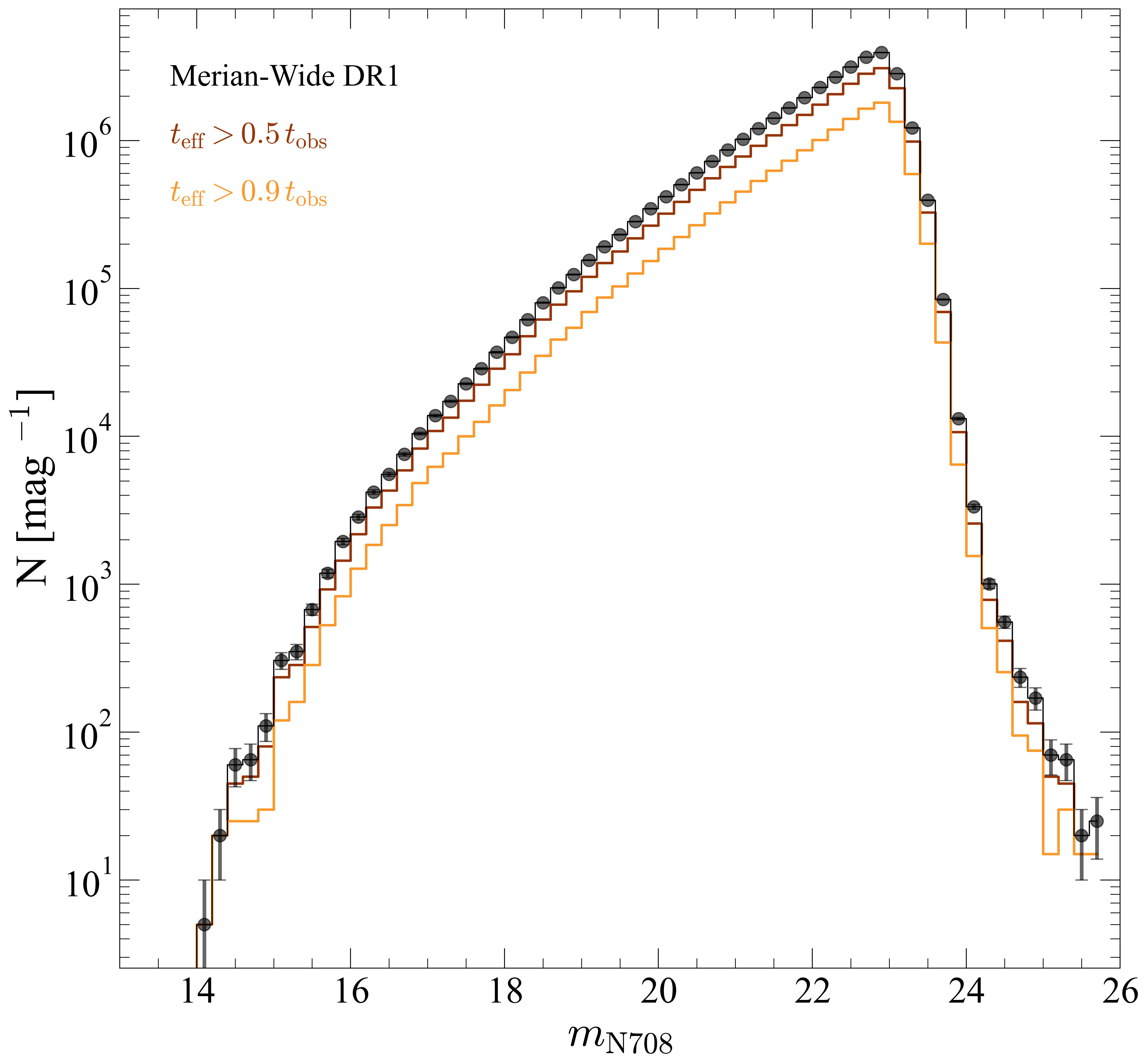}
  \caption{Number counts of objects in the Merian-Wide DR1 photometric catalog (\texttt{SciUse}=1) as a function of the total $m_{N708}$ magnitude, without any correction for incompleteness. The total number counts with no $t_\mathrm{eff}$ cuts is shown in black with Poisson errors. The brown and orange histograms show objects with $t_\mathrm{eff}>0.5\,t_\mathrm{obs}$ and $t_\mathrm{eff}>0.9\,t_\mathrm{obs}$, respectively, where $t_\mathrm{obs}=600\,\mathrm{sec}$ for the $N708$ band.}
  \label{fig:dr_counts}
\end{figure*}

\begin{figure*}[t]
 \centering
 \includegraphics[width=1\textwidth]{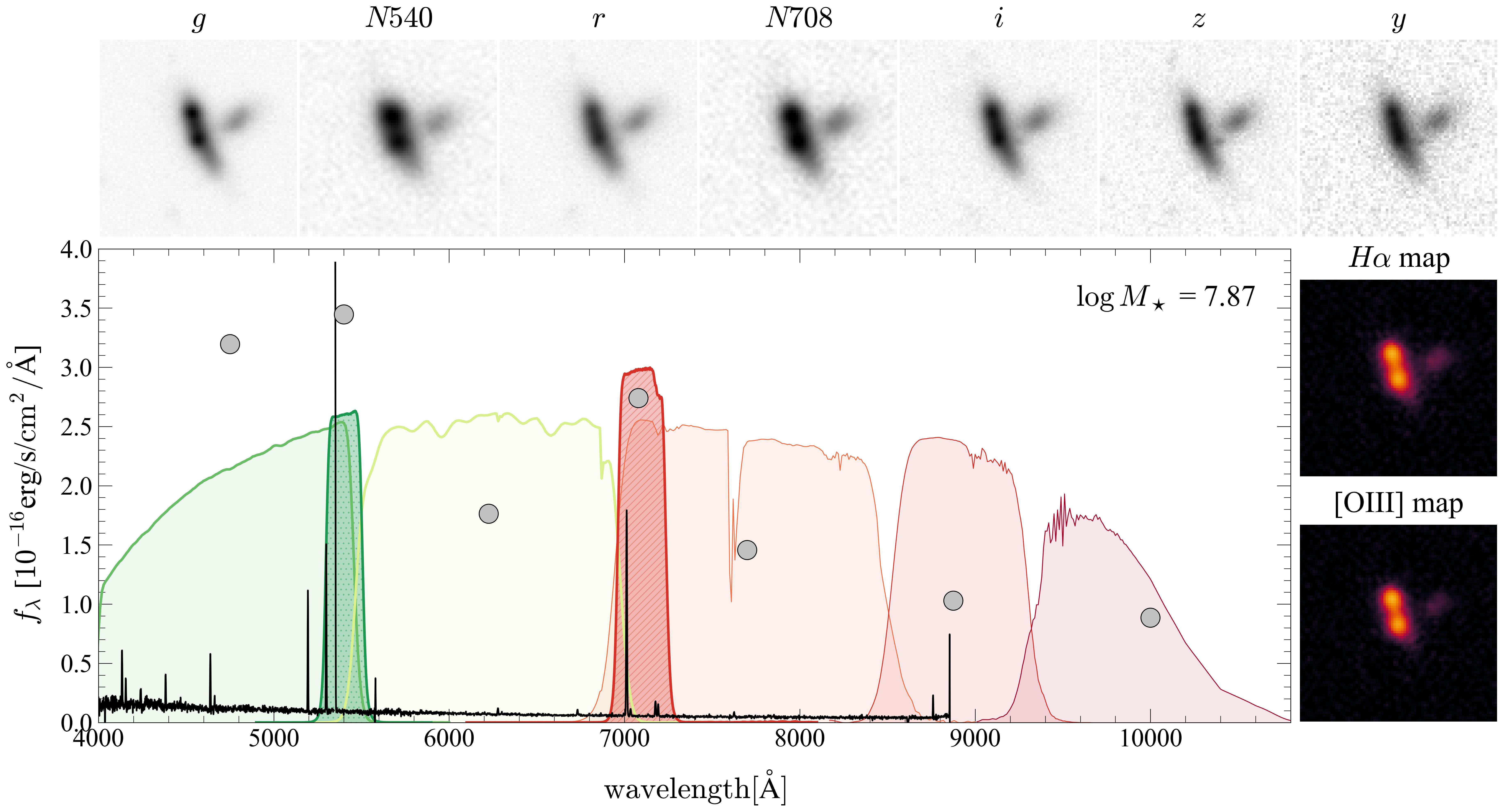}
  \caption{Example dwarf galaxy ($\log(M_\star/M_\odot)=7.87$) from Merian DR1. The main panel shows the galaxy's seven-band photometry (gray symbols) and the GAMA spectrum (black). The top panels (grayscale) show the HSC and Merian cutouts and the bottom right panels show the \Ha\ and \Oiii\ maps generated using the method described in \citet{Mintz:2024}. 
  }
  \label{fig:example_single}
\end{figure*}

\begin{figure*}[t]
 \centering
 \includegraphics[width=0.96\textwidth]{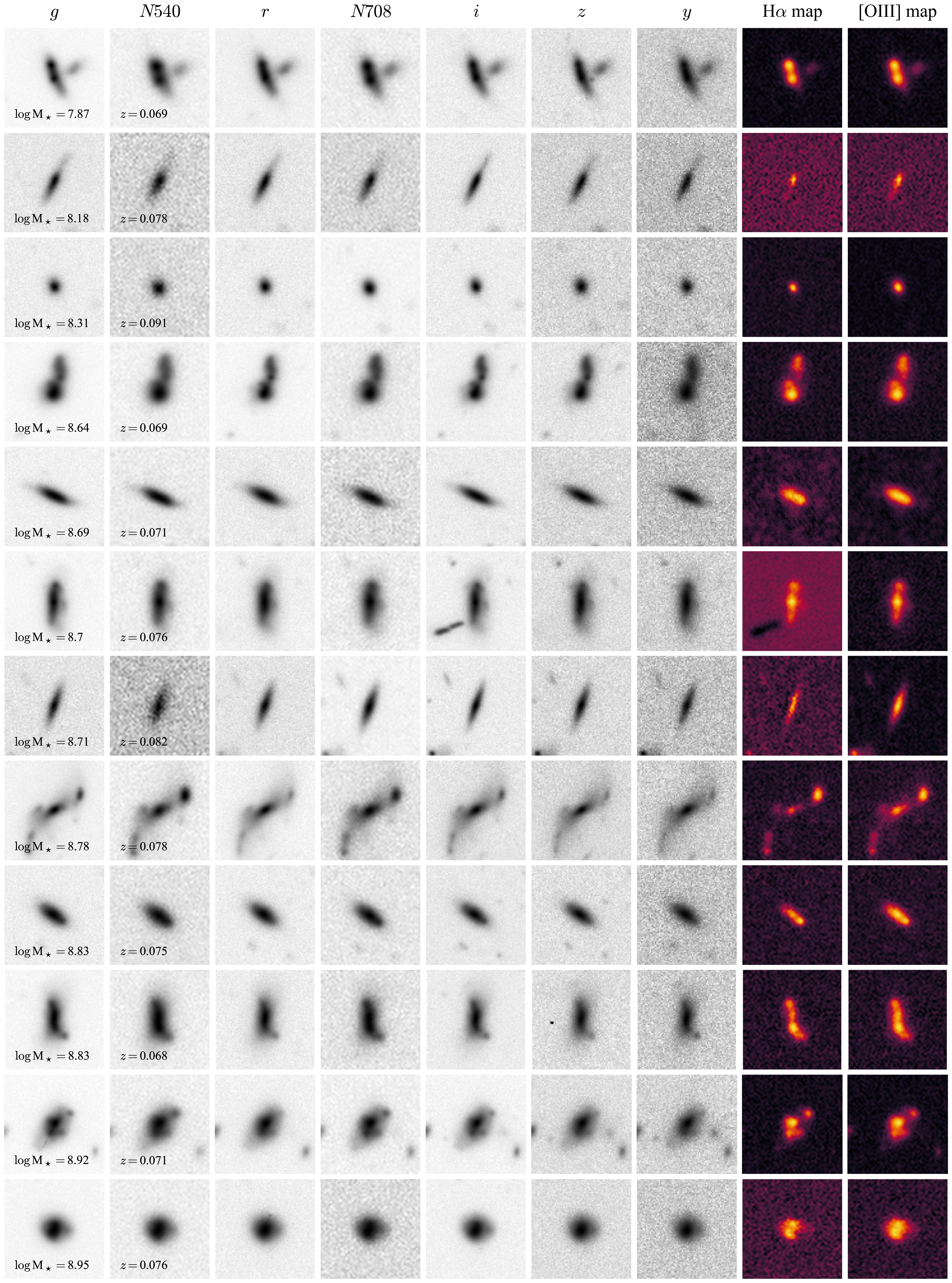}
 \caption{\comment{Postage stamps of 12 dwarf galaxies from the Merian catalog are shown in the seven leftmost columns, corresponding to $g$, $N540$, $r$, $N708$, $i$, $z$, and $y$ filters. Each galaxy has a spectroscopic redshift, obtained from either existing surveys (SDSS and GAMA) or new observations using Magellan/IMACS and Keck/DEIMOS (\S \ref{sec:merian_spec}). The two rightmost columns show the continuum-subtracted \Ha\ and \Oiii\ maps, derived from $N708$ and $N540$, with the continuum estimated via a geometric average of the adjacent broad-band filter fluxes.}}
  \label{fig:gallery}
\end{figure*}

The Merian DR1 photometric catalog comprises 242 columns in total. Key column headers and their descriptions are listed in Table \ref{table:catalog} in Appendix \ref{appendix:app_catalog}. The photometric catalog contains approximately 80 million unique objects across 335 tracts, covering an area of $234\,\mathrm{deg}^2$. By utilizing the \texttt{HEALPix} mask, users can apply a $t_\mathrm{eff}$ cut based on their specific scientific needs, thereby including sources from regions with varying depths. Figure \ref{fig:dr_counts} shows the number counts of all of the objects in the Merian-Wide DR1 catalog as a function of the total $m_{N708}$ magnitude (black) and after applying two exemplary $t_\mathrm{eff}$ cuts. The source detection sensitivity is consistent among the three samples.

\section{Key Science Objectives of the Survey}\label{sec:science_objs}

The Merian Survey described in this manuscript is well-suited for investigating the cosmological and galaxy evolution properties of bright dwarf galaxies. This is achieved by obtaining photometric redshifts for a comprehensive and well-understood sample comprising $\sim 10^5$ galaxies at $0.06<z<0.10$. In the following discussion, we outline several key scientific problems that require a wide range of datasets that Merian has made available. We emphasize the effectiveness of incorporating narrow and medium-band filters alongside broad-band filters for sky imaging, as demonstrated by Merian. This approach showcases the synergistic potential of these combined filters in uncovering astrophysical insights across a broad spectrum of phenomena.

\subsection{What is the dark matter density profile out to the virial radius in dwarf galaxies?}

The distribution and overall dark matter content in dwarf galaxies have far-reaching implications for constraining the nature of dark matter and uncovering the interplay between dark matter particles of various models and baryonic physics at kpc scales \citep{Bullock:2017, Simon:2019, Sales:2022}. Typically, constraints on the dark matter halos hosting dwarf galaxies rely on their internal star and gas kinematics, which primarily access the inner regions of these halos. These baryon-dominated regions probe only a small fraction of their halo virial radii—about 10-20 times smaller than the virial radius and covering less than 1\% of the total volume. Thus, estimating halo mass via star and gas kinematics necessitates extrapolations contingent upon assumptions about the shape of the dark matter profile extending to large, uncharted radii \citep[e.g.][]{Buckley:2018}. 

To further disentangle baryonic effects from dark matter properties, direct observations of dwarf galaxies’ mass profiles are required. The Merian bright dwarf galaxy sample will be sufficiently large to enable direct measurement of dwarfs' halo masses out to the halo virial radius through weak gravitational lensing \citep{Leauthaud:2020, Thornton:2023}. In particular, galaxy-galaxy lensing with the Merian sample of $\sim 10^5$ dwarf galaxies will measure the full halo mass profile through the average weak lensing distortion from background source galaxies with the three-dimensional positions of the Merian sample of foreground lens galaxies.

\subsection{How do dwarf galaxies assemble their mass?}

The build-up of galaxies' stellar masses is the outcome of an intricate interplay of physical phenomena such as gravity, gas cooling and condensation, galaxy-galaxy mergers, and a diverse array of feedback mechanisms \citep[e.g.,][]{Conselice:2014, Naab:2017}. Despite extensive observational and theoretical studies, the specifics and relative contributions of each of these processes in the low-mass regime remain highly uncertain. A key tool for understanding the statistical properties of galaxies is the galaxy stellar mass function (GSMF). The GSMF provides crucial insights into the abundance of cold molecular gas across cosmic epochs which fuel star formation activity and the overall growth of stellar material in the universe \citep[e.g.,][]{Cole:2001, Bell:2003, Bernardi:2013, Driver:2022, Weaver:2023}. Incomplete photometric and spectroscopic samples suggest an intriguing indication of a steepening of the GSMF below $M_\star \sim 10^9\,M_\odot$ \citep{Baldry:2012, Wright:2017}. If confirmed, this upturn could support recent solutions that were proposed for solving the long-standing tension between the present-day stellar mass density with the integral of the cosmic star formation history \citep{Baldry:2003, Wilkins:2008, Leja:2020}.

With its large accessible volume and improved surface brightness sensitivity, Merian is poised to accurately measure the GSMF down to the survey's completeness limit ($M_\star=10^8\,M_\odot$) out to $z = 0.1$. While Merian will achieve a $90\%$ completeness level, it may overlook low-mass galaxies not actively forming stars; however, employing cross-correlation techniques will facilitate the measurement of number densities and properties of these missing galaxies. The low-mass end of the GSMF can then be compared with prior empirical measurements and utilized to test predictions of hydrodynamical simulations incorporating different galaxy formation prescriptions.

\subsection{Explore the connection between star formation and feedback in dwarfs}

Baryonic feedback, and in particular that resulting from star formation, is thought to play a significant role in dwarf galaxy evolution \citep[see, e.g.,][]{Pontzen:2012, Collins:2022}. Although it is essential for reproducing present-day dwarfs, significant challenges and ongoing debates remain regarding the complex interplay between star formation and feedback and its accurate implementation in galaxy formation simulations \citep{Agertz:2015, Sales:2022}. The Merian sample serves as a crucial benchmark for testing prescriptions for star formation and feedback in simulations. Utilizing this sample allows for the measurement of dwarf galaxy distribution across the mass--size plane, considering variables such as environment, star formation rates, and the spatial distribution of star formation (whether concentrated or diffuse). Beyond sizes, diagnostics of the intrinsic shapes \citep{Kado-Fong:2020b, Kado-Fong:2022} can be correlated with central surface brightness, and the environment of galaxies on Mpc scales. In \citet{Mintz:2024}, preliminary findings from a non-parametric morphological characterization of the continuum and \Ha\ emission from the Merian DR1 are presented. Specific star formation rates (SSFR) are shown to increase with the asymmetry of the stellar continuum and the \Ha\ emission; the least active galaxies with $M_\star\lesssim 10^9 M_\odot$ are puffy and diffuse, while those with the highest SSFR have \Ha\ emission that is consistently heterogeneous and compact. Indeed, one of Merian's strengths will lie in its ability to detect and characterize the population of extreme emission line galaxies (EELGs), i.e., highly starbursting galaxies characterized by their strong line emission, and to enable new insights into star formation under extreme conditions.

\subsection{Ancillary Science Goals}

By combining the HSC deep broad-band imaging and the two medium-band data ($N540$ and $N708$), we expect that this dataset will be appealing to a wide range of exciting science questions. Merian will provide unique data for identifying and studying extended narrow-line regions (ENLR) in active galaxies in WISE-selected AGN at $z\sim 0.4$ and $z\sim 0.9$ \citep[e.g.,][]{Liu:2013}. Merian will also identify the largest sample (hundreds) of Enormous \Lya\ Nebulae (ELANe, \citealt{Cai:2017}) at $z=3.3$ and $z=4.8$.

\section{Summary}\label{sec:summary}

In this manuscript, we present the Merian survey, a large program conducted with the Blanco/DECam that obtained wide-field imaging over $\sim 750\,\mathrm{deg}^2$ using two custom-built medium-band filters centered at $\lambda_\mathrm{c}=7080\Angs$ ($\Delta\lambda=275\Angs$) and $\lambda_\mathrm{c} = 5400\Angs$ ($\Delta\lambda = 210\Angs$). Combined with deep broad-band imaging ($grizy$) from the HSC-SSP survey, Merian provides photometric redshifts and seven-band deep imaging to $\sim 10^5$ dwarf galaxies ($M_\star=10^8-10^9\,M_\odot$) in well-studied extragalactic equatorial survey fields (GAMA, COSMOS, XMM-LSS, VVDS, and SXDS). 
Merian's capability to capture the redshifted \Ha\ and \Oiii\ emission within the $z=0.06-0.10$ redshift window enables the measurement of photometric redshifts for all star-forming galaxies within its survey footprint. The photometric redshifts enable weak lensing measurements of dwarfs down to $M_\star=10^8\,M_\odot$, and facilitate a multitude of studies investigating the physical properties of dwarfs through their continuum and emission line characteristics.

\comment{This manuscript also presents the first data release (DR1), which covers $234\,\mathrm{deg}^2$ of full color-full depth (FCFD) imaging from a total DR1 full-color survey area of $320\,\mathrm{deg}^2$. The final survey footprint spans $584\,\mathrm{deg}^2$ of FCFD from a total full-color coverage of $733\,\mathrm{deg}^2$. For DR1, we have adapted the LSST Science Pipelines code to perform the reduction of the DECam data with coadds resampled to a pixel grid common to the new DECam images and the HSC-SSP images. The reduction pipeline employs improvements in the astrometric and photometric calibration methods, artifact rejection, and sky subtraction scheme as described in \citet{Aihara:2022} and in source separation (``deblending'') as described in \citet{Melchior2018}. Of particular importance is the utilization of the Gaussian Aperture and PSF (GAaP) Photometry \citep{Kuijken:2008} for measuring accurate galaxy colors from matched-aperture fluxes, resulting in improved photometric redshifts (presented in an accompanying manuscript; Luo et al. 2024, in prep). The final data release will encompass the entire survey area, adhering to the same survey strategy but incorporating a more recent release of the LSST Science Pipelines for improved data processing and catalog generation.}

Merian's uniqueness lies in its utilization of existing data from the deep wide-area HSC-SSP imaging survey, coupled with the widest-area imaging survey conducted with narrow-medium band filters. This approach not only enhances the power of broad-band imaging surveys by providing significantly improved photometric redshifts but also by unveiling crucial insights into star formation and other physical processes through the analysis of spatially-resolved \Ha\ and \Oiii\ emission maps. With a uniform and well-understood selection function, Merian's strategy offers a novel and cost-effective way to obtain essential details such as redshifts, stellar masses, and star formation rates, given the prohibitive nature of obtaining spectra for all galaxies down to such low masses across a wide area. Looking ahead, with the emergence of new wide-field imaging surveys such as Euclid, Rubin/LSST, and Roman, this approach holds promise for obtaining complementary data through narrow or medium-band imaging. It facilitates the maximization of data from these surveys and ensures a homogeneous selection function of targets for more resource-intensive spectroscopic surveys.

\acknowledgments

We are grateful to Dustin Lang, David Schlegel, and Eddie Schlafly for their invaluable assistance during the early stages of survey planning. We also appreciate Dustin Lang’s efforts in integrating Merian data into the LegacyViewer, which greatly facilitated our data quality checks. We thank Konrad Kuijken for his support with GaaP. This project would not have been possible without the exceptional support and dedication of the NOIRLab scientific staff, especially Guillermo Damke, Clara Martinez-Vazquez, and Alistair Walker.

S.D. acknowledges generous support from a Carnegie-Princeton Fellowship, through Princeton University and the Carnegie Observatories and supported provided by NASA through Hubble Fellowship grant HST-HF2-51454.001-A awarded by the Space Telescope Science Institute, which is operated by the Association of Universities for Research in Astronomy, Incorporated, under NASA contract NAS5-26555. AK was supported in part by the National Science Foundation through Cooperative Agreement 1258333 managed by the Association of Universities for Research in Astronomy (AURA), and the Department of Energy under Contract No. DE-AC02-76SF00515 with the SLAC National Accelerator Laboratory. GEM acknowledges support from the University of Toronto Arts \& Science Post-doctoral Fellowship program, the Dunlap Institute, and the Natural Sciences and Engineering Research Council of Canada (NSERC) through grant RGPIN-2022-04794. JIR would like to acknowledge support from STFC grants ST/Y002865/1 and ST/Y002857/1.

This material is based upon work supported by the National Science Foundation under Grant No. 2106839. This project used data obtained with the Dark Energy Camera (DECam), which was constructed by the Dark Energy Survey (DES) collaboration. Funding for the DES Projects has been provided by the US Department of Energy, the US National Science Foundation, the Ministry of Science and Education of Spain, the Science and Technology Facilities Council of the United Kingdom, the Higher Education Funding Council for England, the National Center for Supercomputing Applications at the University of Illinois at Urbana-Champaign, the Kavli Institute for Cosmological Physics at the University of Chicago, Center for Cosmology and Astro-Particle Physics at the Ohio State University, the Mitchell Institute for Fundamental Physics and Astronomy at Texas A\&M University, Financiadora de Estudos e Projetos, Fundação Carlos Chagas Filho de Amparo à Pesquisa do Estado do Rio de Janeiro, Conselho Nacional de Desenvolvimento Científico e Tecnológico and the Ministério da Ciência, Tecnologia e Inovação, the Deutsche Forschungsgemeinschaft and the Collaborating Institutions in the Dark Energy Survey.

The Collaborating Institutions are Argonne National Laboratory, the University of California at Santa Cruz, the University of Cambridge, Centro de Investigaciones Enérgeticas, Medioambientales y Tecnológicas–Madrid, the University of Chicago, University College London, the DES-Brazil Consortium, the University of Edinburgh, the Eidgenössische Technische Hochschule (ETH) Zürich, Fermi National Accelerator Laboratory, the University of Illinois at Urbana-Champaign, the Institut de Ciències de l’Espai (IEEC/CSIC), the Institut de Física d’Altes Energies, Lawrence Berkeley National Laboratory, the Ludwig-Maximilians Universität München and the associated Excellence Cluster Universe, the University of Michigan, NSF’s NOIRLab, the University of Nottingham, the Ohio State University, the OzDES Membership Consortium, the University of Pennsylvania, the University of Portsmouth, SLAC National Accelerator Laboratory, Stanford University, the University of Sussex, and Texas A\&M University.

Based on observations at Cerro Tololo Inter-American Observatory, NSF’s NOIRLab (NOIRLab Prop. ID 2020B-0288; PI: A. Leauthaud), which is managed by the Association of Universities for Research in Astronomy (AURA) under a cooperative agreement with the National Science Foundation.

The Hyper Suprime-Cam (HSC) collaboration includes the astronomical communities of Japan and Taiwan and Princeton University. The HSC instrumentation and software were developed by the National Astronomical Observatory of Japan (NAOJ), the Kavli Institute for the Physics and Mathematics of the Universe (Kavli IPMU), the University of Tokyo, the High Energy Accelerator Research Organization (KEK), the Academia Sinica Institute for Astronomy and Astrophysics in Taiwan (ASIAA), and Princeton University. Funding was contributed by the FIRST program from the Japanese Cabinet Office, the Ministry of Education, Culture, Sports, Science and Technology (MEXT), the Japan Society for the Promotion of Science (JSPS), Japan Science and Technology Agency (JST), the Toray Science Foundation, NAOJ, Kavli IPMU, KEK, ASIAA, and Princeton University. 

This paper makes use of software developed for Vera C. Rubin Observatory. We thank the Rubin Observatory for making their code available as free software at http://pipelines.lsst.io/.

This paper is based on data collected at the Subaru Telescope and retrieved from the HSC data archive system, which is operated by the Subaru Telescope and Astronomy Data Center (ADC) at NAOJ. Data analysis was in part carried out with the cooperation of Center for Computational Astrophysics (CfCA), NAOJ. We are honored and grateful for the opportunity of observing the Universe from Maunakea, which has the cultural, historical and natural significance in Hawaii.

The authors are pleased to acknowledge that the work reported in this paper was substantially performed using the Princeton Research Computing resources at Princeton University, a consortium of groups led by the Princeton Institute for Computational Science and Engineering (PICSciE) and the Office of Information Technology's Research Computing.

\appendix
\section{Merian DR1 Photometric Catalog Format}\label{appendix:app_catalog}

The full Merian photometric catalog includes 242 columns utilizing object information and measurements from the Merian and HSC-SSP S20A surveys. Selected columns are presented in Table \ref{table:catalog}. X denotes a filter name spanning $N540$, $N708$, $grizy$, unless noted otherwise in the column description. Y denotes the aperture size used in the GAaP photometry (e.g., 1p0, 1p5, etc.).

\begin{table}[ht]{}
\setlength{\tabcolsep}{20pt}
\renewcommand{\arraystretch}{1.2}
\caption{Photometeric Catalog Columns}
\centering 
\begin{tabular}{p{5cm} p{9cm} p{0.3cm}} \hline\hline 
Column name & Description & Source \\ [0.5ex]
\hline 
\texttt{objectId\_Merian} & Unique object identifier & Merian \\
\hline 
\texttt{coord\_ra\_Merian} & ICRS right ascension of object centroid & '' \\
\hline 
\texttt{coord\_dec\_Merian} & ICRS declination of object centroid & '' \\
\hline 
\texttt{ebv\_Merian} & Galactic reddening & '' \\
\hline 
\texttt{tract\_Merian} & Skymap tract ID & '' \\
\hline 
\texttt{patch\_Merian} & Skymap patch ID & '' \\
\hline 
\texttt{detect\_isPrimary\_Merian} & True if object seed has no children \& is in the inner region of a coadd patch \& is in the inner region of a coadd tract \& is not a sky object & '' \\
\hline 
\texttt{X\_psfFlux\_Merian} & X-band flux from linear least-squares fit of PSF model  & '' \\
\hline 
\texttt{X\_psfFluxErr\_Merian} & X-band flux uncertainty from linear least-squares fit of PSF model  & '' \\
\hline 
\texttt{X\_gaapYFlux\_Merian} & GAaP flux with Y aperture for X-band & '' \\
\hline 
\texttt{X\_gaapYFlux\_Merian} & GAaP flux uncertainty with Y aperture for X-band  & '' \\
\hline 
\texttt{X\_gaapYFlux\_aperCorr\_Merian} & X-band GAaP aperture corrected flux with Y aperture ($grizy$) & '' \\
\hline 
\texttt{X\_extendedness\_Merian} & Set to 1 for extended sources, 0 for point sources & '' \\
\hline 
\texttt{X\_blendedness\_Merian} & Measure of how much the flux is affected by neighbors & '' \\
\hline 
\texttt{X\_cModelFlux\_Merian} & X-band flux from the final cModel fit & '' \\
\hline 
\texttt{X\_cModelFluxErr\_Merian} & X-band flux uncertainty from the final cModel fit  & '' \\
\hline 
\texttt{X\_inputCount\_Merian} & Number of images contributing at the center  & '' \\
\hline 
\texttt{X\_cModel\_flag\_Merian} & Flag set if the final cModel fit  & '' \\
\hline 
\texttt{X\_pixelFlags\_bad\_Merian} & Bad pixel in the Source footprint  & '' \\
\hline 
\texttt{X\_pixelFlags\_clippedCenter\_Merian} & Source center is close to CLIPPED pixels  & '' \\
\hline 
\texttt{X\_pixelFlags\_cr\_Merian} & Cosmic ray in the Source footprint & '' \\
\hline 
\texttt{X\_pixelFlags\_crCenter\_Merian} & Cosmic ray in the Source center & '' \\
\hline 
\texttt{X\_pixelFlags\_edge\_Merian} & Source is outside usable exposure region & '' \\
\hline 
\texttt{X\_pixelFlags\_interpolated\_Merian} & Interpolated pixel in the Source footprint & '' \\
\hline 
\texttt{X\_pixelFlags\_interpolatedCenter\_Merian} & Interpolated pixel in the Source center & '' \\
\hline 
\texttt{X\_pixelFlags\_saturated\_Merian} & Saturated pixel in the Source footprint & '' \\
\hline 
\texttt{X\_pixelFlags\_suspect\_Merian} & Source's footprint includes suspect pixels & '' \\
\hline 
\texttt{X\_pixelFlags\_suspectCenter\_Merian} & Source's center is close to suspect pixels & '' \\
\hline 
\texttt{X\_centroid\_flag\_Merian} & General failure flag & '' \\
\hline 
\texttt{IsMask\_Merian} & Set to 1 if the object is masked, 0 for unmasked objects & '' \\
\hline 
\texttt{ra\_HSCS20A} & ICRS right ascension of object centroid & HSC-SSP S20A \\
\hline 
\texttt{dec\_HSCS20A} & ICRS declination of object centroid & '' \\
\hline 
\texttt{X\_extendedness\_value\_HSCS20A} & Set to 1 for extended sources, 0 for point sources & '' \\
\hline 
\texttt{X\_blendedness\_abs\_HSCS20A} & Measure of how much the flux is affected by neighbors & '' \\
\hline 
\texttt{hsc\_match} & Set to 1 if a match to HSC-SSP S20A found, 0 otherwise & -- \\
\hline 
\texttt{PhotUse} & Set to 1 when a source is not masked by the bright star mask, 0 otherwise & -- \\
\hline 
\texttt{SciUse} & Set to 1 if a source is not masked, has a full-color full-depth, likely not a star, fulfill several quality criteria (see Section \ref{sec:photo_cat}) & -- \\
\hline 
\hline 
\end{tabular}
\label{table:catalog}
\end{table}


\end{document}